\documentclass[12pt,aps,prd,floatfix,nofootinbib,a4paper,nosuperscriptaddress]{revtex4-2}
\pdfoutput=1

\usepackage{amsmath, amssymb, amsfonts, amsthm, latexsym, epsfig, mathrsfs, xcolor, bbm, slashed, braket, thmtools}

\usepackage[all]{xy}

\usepackage[inline]{enumitem}

\usepackage{setspace}
\usepackage[marginal, multiple]{footmisc}

\usepackage[T1]{fontenc}
\usepackage[utf8]{inputenc}
\usepackage{lmodern}

\usepackage{comment}

\usepackage[colorlinks, allcolors=blue!70!black, linktocpage, hypertexnames=false]{hyperref}

\definecolor{indigo(dye)}{rgb}{0.0, 0.25, 0.42}

\numberwithin{equation}{section}

\usepackage{cleveref}

\usepackage{microtype}

\usepackage{floatrow}
\let\OLDtableofcontents\tableofcontents
\renewcommand\tableofcontents[1]{%
    {\baselineskip 0.5ex %
	\OLDtableofcontents{#1}}%
}

\let\OLDthebibliography\thebibliography
\renewcommand\thebibliography[1]{%
	\setstretch{1.079} % 1/2 goldenratio
	\OLDthebibliography{#1}%
	\small %
	\setlength{\itemsep}{0.2\baselineskip} % goldenratio for separating items %
}

\let\OLDfootnote\footnote
\renewcommand\footnote[1]{%
	\setlength{\footnotesep}{0.75\baselineskip}%
	{\footnotesize \OLDfootnote{#1}}%
}

\setlist[enumerate]{noitemsep, label=(\arabic*), ref=(\arabic*)}

\newlist{condlist}{enumerate}{2}
\setlist[condlist,1]{noitemsep, topsep=0pt, label=(\arabic*), ref=(\arabic*)}
\setlist[condlist,2]{noitemsep, label=(\alph*), ref=(\arabic{condlisti}.\alph*)}
\crefname{condlisti}{condition}{conditions}
\crefname{condlistii}{condition}{conditions}

\newlist{propertylist}{enumerate}{1}
\setlist[propertylist,1]{noitemsep, topsep=0pt, label=(\arabic*), ref=(\arabic*)}
\crefname{propertylisti}{Property}{Properties}

\renewcommand\thesection{\arabic{section}}
\renewcommand\thesubsection{\arabic{subsection}}

\makeatletter
\def\p@subsection{\thesection.}
\def\p@subsubsection{\thesection.\thesubsection.}
\makeatother 

\theoremstyle{plain}

\theoremstyle{definition}

\declaretheorem[style=remark,qed=$\scriptstyle{\blacksquare}$,numberwithin=section]{remark} 

\creflabelformat{equation}{#2#1#3}

\crefname{section}{\S}{\S}
\crefname{appendix}{appendix}{appendices}
\crefname{figure}{Fig.}{Figs.}
\crefname{table}{Table}{Tables}

\crefname{definition}{Def.}{Defs.}
\crefname{prop}{Prop.}{Props.}
\crefname{lemma}{Lemma}{Lemmas}
\crefname{corollary}{Cor.}{Cors.}
\crefname{thm}{Theorem}{Theorems}
\crefname{remark}{Remark}{Remarks}

\crefname{ass}{Assumptions}{Assumptions}
\crefname{property}{Properties}{Properties}

\newcommand{\be}{\begin{equation}\begin{aligned}}
\newcommand{\ee}{\end{aligned}\end{equation}}

\newcommand{\lb}{\left}
\newcommand{\rb}{\right}

\newcommand{\mc}{\mathcal}

\newcommand{\bb}{\mathbb}

\newcommand{\eqsp}{\, ,\quad} %shrtct for space in eqns

\newcommand{\defn}{\mathrel{\mathop:}=} %shrtct for definition operator

\newcommand{\norm}[1]{\lb\Vert\, #1 \,\rb\Vert}		% norm
\newcommand{\op}[1]{\boldsymbol{#1}}
\renewcommand{\1}{\op{1}}

\newcommand{\Dop}[2]{\ket{#1}\!\bra{#2}}

\newcommand{\Tord}{{\mc T}}

\newcommand{\inn}{\textrm{in}}

\newcommand{\out}{\textrm{out}}

\newcommand{\free}{\textrm{free}}

\let\oldsetminus\setminus
\renewcommand{\setminus}{\!\oldsetminus\!} 

\newcommand{\pd}[2][]{\frac{\partial #1}{\partial #2}} 

\newcommand{\td}[2][]{\frac{d #1}{d #2}} 

\let\oldint\int
\renewcommand{\int}{\oldint\limits}

\let\oldlim\lim
\renewcommand{\lim}{\oldlim\limits}

\renewcommand{\bar}{\overline}

\newcommand{\Hilb}{\mathscr{H}}

\newcommand{\antiHilb}{%
\hspace{4pt} % puts a leading space which Latex skips with below code
  \vbox{%
    \hrule height 0.5pt%                  % Line above with certain width
    \kern0.25ex%                          % Distance between line and content
    \hbox{%
      \kern-0.3em%                        % Distance between content and left side of box, negative values for lines shorter than content
      \ifmmode\Hilb\else\ensuremath{\Hilb}\fi%  % The content, typeset in dependence of mode
      \kern0em%                        % Distance between content and right side of box, negative values for lines shorter than content
    }% end of hbox
  }% end of vbox
}

\newcommand{\nfrac}[2]{{{}^#1\!\!/\!_#2}}

\newcommand{\half}{\nfrac{1}{2}}

\DeclareMathOperator{\sgn}{\text{\rm sgn}}

\DeclareMathOperator{\pv}{\text{\rm pv}}

\DeclareMathOperator{\Pf}{\text{\rm Pf}}

\renewcommand{\Re}{{\rm Re\,}}

\newcommand{\abs}[1]{\lb\vert\, #1 \,\rb\vert}

\DeclareMathOperator*{\slim}{s-lim}
\let\oldslim\slim
\renewcommand{\slim}{\oldslim\limits}

\DeclareMathOperator*{\wlim}{w-lim}
\let\oldwlim\wlim
\renewcommand{\wlim}{\oldwlim\limits}

\begin{document}

\setstretch{1.2}

% Title Page

\title{Infrared finite scattering theory: Non-relativistic Coulomb scattering}
% Authors
\author{Kartik Prabhu}
\email{kartikprabhu@rri.res.in}
\affiliation{Raman Research Institute, Sadashivanagar, Bengaluru 560080, India.}

\author{Shivam Upadhyay}
\email{shivamu@rrimail.res.in}
\affiliation{Raman Research Institute, Sadashivanagar, Bengaluru 560080, India.}

\begin{abstract}
We revisit non-relativistic quantum-mechanical scattering in a Coulomb potential. The long-range nature of the interaction leads to divergent integrals in conventional S-matrix calculations using the Born series. Dollard showed that a well-defined scattering operator can be constructed by replacing the free Hamiltonian \(H_0\) with an appropriate asymptotic Hamiltonian \(H_{\rm D}\). However, explicit perturbative calculations within Dollard's framework have remained scarce. We develop a perturbative formulation of the Dollard scattering operator in which the Coulomb divergences cancel to all orders. Perturbatively, we demonstrate this cancellation explicitly through second Born order and show that the resulting amplitude reproduces the known exact distributional Coulomb amplitude, including its forward-scattering contribution. With the aim of generalizing our analysis to quantum field theory, we reformulate the problem in the Heisenberg picture. We construct a new set of canonical variables through a time-dependent canonical transformation that removes the logarithmic Coulomb correction from the asymptotic phase-space coordinates. Using these variables, we construct the corresponding Heisenberg S-matrix through several complementary approaches. This leads to a formulation of the Lehmann--Symanzik--Zimmermann (LSZ) reduction formalism adapted to the modified asymptotic dynamics, which we relate to the Dollard scattering operator. This construction provides a framework for extending the LSZ formalism to infrared divergences in relativistic quantum field theories with long-range interactions and massless fields.
\end{abstract}

\maketitle
\newpage 
\tableofcontents

%%======================================================
\section{Introduction}
\label{sec:intro}

Infrared (IR) divergences are a characteristic feature of quantum field theories with long-range interactions, most notably quantum electrodynamics (QED) and quantum gravity. They arise from the interaction of scattering massive fields with massless quantum fields, which can carry arbitrarily low energy due to the absence of a mass gap. Generic scattering processes produce an ``infinite number'' of these low-energy ``soft particles'' relative to the usual Minkowski invariant vacuum state. Thus, the asymptotic state cannot be represented by an ordinary Fock space state. The resulting IR divergences are not simply a technical complication of perturbation theory; they signal a mismatch between the asymptotic states used in the conventional scattering formalism and the actual long-range structure of the theory.

This problem has been understood in several complementary ways. Early attempts to construct an appropriate scattering theory for QED include the work of Chung \cite{Chung:1965zza} and Kibble \cite{Kibble:1968sfb,Kibble:1968oug,Kibble:1968npb,Kibble:1968lka}, culminating in the construction of dressed asymptotic states by Faddeev and Kulish \cite{Kulish:1970ut}. These states incorporate the long-range electromagnetic field associated with charged particles and modify the asymptotic dynamics entering the definition of the scattering operator. A related perspective emerged from the construction of inequivalent representations of the asymptotic massless fields, the \emph{memory Hilbert spaces} \cite{Ashtekar:1987tt,Ashtekar:2018lor,Prabhu:2022zcr}. From this viewpoint, the difficulty with the conventional Fock space description can be seen as the failure to accommodate scattering states with different asymptotic electromagnetic fields within a single separable Hilbert space representation.\footnote{We note that the work of Kibble \cite{Kibble:1968sfb} uses a ``very large'' \emph{non-separable} Hilbert space to accommodate scattering states, which makes the computations quite cumbersome.} This viewpoint is in line with the recent ideas connecting the infrared problem to the asymptotic symmetries and conservation laws of gauge theories and gravity. In particular, the relation between dressed states and asymptotic symmetries has been developed in several papers \cite{Kapec:2017tkm,Choi:2017ylo,Gabai:2016kuf,Prabhu:2022zcr}. In this framework, the IR divergence problem has been understood as a consequence of using asymptotic states that do not account for the conservation of charges associated with asymptotic symmetries. For a rigorous treatment of infrared divergences in QED, see \cite{Duch:2019wpf}, and \cite{Hannesdottir:2019opa} for a formulation of the Faddeev--Kulish dressing in terms of soft Wilson lines constructed from a soft collinear effective field theory.

There remains, however, a practical challenge. Although constructions such as that of Faddeev and Kulish provide a prescription for modifying the asymptotic dynamics, implementing this prescription in perturbative calculations of scattering amplitudes is often cumbersome. There are only a handful of explicit calculations of IR-finite amplitudes; see, for example, \cite{Forde:2003jt,Hannesdottir:2019opa}. In particular, it is desirable to have a formulation of an IR-finite S-matrix which can be computed perturbatively and in which the cancellation of IR divergences can be exhibited directly, without requiring the construction of complicated dressed states. Closely related is the question of how to formulate Lehmann--Symanzik--Zimmermann (LSZ) reduction for theories with long-range interactions where the asymptotic dynamics are not free. Various attempts to formulate LSZ-type constructions in the presence of long-range interactions have been made \cite{Zwanziger:1973if,Papanicolaou:1976sv,Dybalski:2017mip,Lippstreu:2023vvg,Lippstreu:2025jit}, but a systematic perturbative framework incorporating the modified asymptotic dynamics remains an important problem. Developing such a framework is the main motivation for the present work.

A simple setting in which these questions can be investigated explicitly is non-relativistic scattering in a Coulomb potential. Despite its simplicity, the Coulomb problem captures the essential difficulty associated with long-range interactions. On the one hand, the Coulomb scattering problem is simple enough that one can find the exact energy eigenstates and the S-matrix \cite{Landau:Lifschitz, Lippstreu:2023vvg} (see also \cref{sec:Coulomb-review}); but these exact methods cannot be directly applied to more general cases.\footnote{For instance, it is well-known that the Coulomb scattering problem has conserved \emph{Runge-Lenz vector} in addition to the conserved energy and angular momentum \cite{Bander:1965im,Zwanziger1967,Lippstreu:2023vvg}. However, this additional conservation law does not hold for any other potential, and so we will not use the Runge-Lenz vector explicitly in our analysis.} On the other hand, the Coulomb S-matrix cannot be formulated perturbatively using the Born-Dyson series. This is because the interacting dynamics do not approach free dynamics at asymptotically early or late times. This is already apparent in the classical theory: for a particle of mass \(m\) and asymptotic momentum \(\vec p_\infty\), the late-time trajectory in a repulsive Coulomb potential \(V(\vec x)=\alpha/|\vec x|\) takes the form \cite{Reed-Simon3}
\be
    \vec x(t) \sim \frac{\vec p_\infty}{m}t
    -\frac{\alpha m\vec p_\infty}{p_\infty^3}
    \ln \abs{4H_0t}
    +\vec x_\infty+O(\ln t/t) \,.
\label{eq:intro-x-falloff}
\ee
Thus, even at arbitrarily large times, the particle's trajectory differs from that of a freely evolving particle by a logarithmic correction. In the quantum theory, this causes the integrals in the conventional Born expansion of the Coulomb scattering amplitude to diverge beyond first order, and the na\"ive series does not capture the fact that amplitude is not an ordinary function in the forward direction.

The structure of these divergences was understood in the early studies of Coulomb scattering. Dalitz \cite{Dalitz:1951ah} conjectured that the divergent terms appearing in higher Born approximations can be collected into a phase multiplying the finite part of the scattering amplitude. Weinberg \cite{Weinberg:1965nx} subsequently proved this exponentiation and showed that it reflects a more general infrared structure of scattering in theories with long-range interactions including QFT. These results indicate that the divergences of the conventional perturbative expansion do not represent a divergence of the physical scattering amplitude itself. Rather, they arise because the conventional scattering prescription uses free asymptotic dynamics in a situation where these dynamics are inadequate.

A systematic resolution of this problem in the Schr\"odinger picture was provided by Dollard \cite{Dollard:1964cok,Dollard-review}. Instead of using the free evolution as a reference at large times, Dollard defined a modified evolution which retains the long-range part of the Coulomb interaction through the Dollard potential \(V_D \sim 1/\abs{t}\) (see \cref{eq:V-Dollard}). Dollard proved the existence of the corresponding modified M\o ller operators and thereby established a well-defined S-matrix operator for the Coulomb scattering. As far as we are aware, the only explicit perturbative implementation of Dollard's construction is due to Papanicolaou \cite{Papanicolaou:1976sv}, which uses the closed-form Coulomb Green's functions \cite{Schwinger:1964zzb}, together with a corresponding modified LSZ reduction formula. Recently, Lippstreu has also carried out perturbative calculations in the relativistic Coulomb problem \cite{Lippstreu:2023vvg,Lippstreu:2025jit}, using a basis of Coulomb scattering states. While these provide a direct treatment of the Coulomb problem, the reliance on exact Coulomb Green's functions and Coulomb wavefunctions makes the approach difficult to generalize to quantum field theories, where such closed-form expressions are generally unavailable.

In this paper, we develop a direct implementation of Dollard's construction into the perturbative Born-Dyson expansion, without requiring Coulomb scattering wavefunctions or Coulomb Green's functions. As explained above, the Born integrals for the Coulomb potential diverge, and so we start with a regularized Coulomb potential. While it is common to use a Yukawa potential as a regulated Coulomb potential \cite{Dalitz:1951ah,Kacser-C,Weinberg:1965nx}, for our computations it will be simpler to use an analytic regulator with the regulated potential \(V_\epsilon \sim \alpha/{\abs{\vec x}^{1+2\epsilon}}\) (see  \cref{eq:FourierV}). For \(\epsilon > 0\), this regulated potential does have a Born series with finite Born integrals, and no Dollard modification is necessary. However, since we are interested in the \(\epsilon \to 0^+\) limit (to retrieve the Coulomb potential), we also include the Dollard modification with a similarly regulated Dollard potential, see \cref{eq:FourierVD}. We then compute the S-matrix \(\op S_\epsilon\) for this regulated potential and Dollard correction. In particular, we show that the regulated Dollard scattering amplitude can be written as the standard Dyson S-matrix along with a phase correction (see \cref{sec:Born-series})
\be
    \braket{p_2 | \op S_\epsilon | p_1}
    =
    \exp\left( i (2 \pi)^\half V_{\rm D,\epsilon}(p,0) \right)
    \braket{p_2 |
    \Tord{\exp\left( - i \int_{-\infty}^{+\infty} dt\,
    \op V_{\rm I,\epsilon}(t) \right)}
    | p_1}.
    \label{eq:intosmatrix}
\ee
Here, $p=p_2=p_1$ by energy conservation and \(V_{\rm I,\epsilon}\) is the regulated Coulomb interaction in the interaction picture. The \(V_{\rm D,\epsilon}(p,0)\) is the zero-frequency component of the regulated Dollard potential (\cref{sec:Dollard}). Although the two factors on the right-hand side are individually divergent as the regulator is removed, their product has a well-defined limit. Using the analysis of Taylor \cite{Taylor:1973da}, we argue that in the limit that the regulator is removed, \(\epsilon \to 0^+\), \(\braket{p_2|\op S_\epsilon|p_1}\) reproduces the known exact Coulomb scattering amplitude.

We also demonstrate the cancellation of divergences explicitly through the second Born order. At first order, the Dollard modified expansion gives (\cref{sec:first-order})
\be
    \braket{p_2 | \op S_\epsilon^{(1)} | p_1}
    =
    -i\left[
    (2\pi)^{-\half} V_\epsilon(\vec p_2-\vec p_1)
    \delta(E_2-E_1)
    -(2\pi)^\half V_{\rm D,\epsilon}(p_2,0)
    \delta(\vec p_2-\vec p_1)
    \right].
\ee
We show that the forward singularity arising from the Fourier transform of the Coulomb potential is cancelled by the Dollard potential contribution and the first-order amplitude is well-defined as a distribution, including in the forward direction (see \cref{eq:firstorderamplitude}). At second order, the expansion gives (\cref{sec:second-order})
\be
    \braket{p_2|\op S_\epsilon^{(2)}|p_1}
    ={}&
    -\pi V_{\rm D,\epsilon}^2(p_2,0)
    \delta(\vec p_2-\vec p_1)
    +V_{\rm D,\epsilon}(p_2,0)\,
V_\epsilon(\vec p_2-\vec p_1)\,
\delta(E_2-E_1)\\
    &+i\frac{2m}{(2\pi)^2}\,
\delta(E_2-E_1)
\int d^3\ell\,
\frac{
V_\epsilon(\vec p_2-\vec\ell)\,
V_\epsilon(\vec\ell-\vec p_1)
}{
\ell^2-p_2^2-i0^+
} \,.
\ee
At this order, the ordinary second Born integral in the second line is divergent, but we show this divergence is cancelled by the corresponding terms from the Dollard potential. Beyond the cancellation of the divergence, the resulting finite expression reproduces the distributional structure of the exact Coulomb scattering amplitude (see \cref{eq:secondorderamplitude}). In particular, our second-order result agrees with the perturbative expansion of the exact distributional amplitude obtained by Herbst \cite{Herbst:1974hi}. This agreement provides a non-trivial check of the construction, since the forward-scattering contribution is essential for recovering the correct distributional S-matrix. Also, note that only the free propagator/Green's function appears in the computations, and we do not need to deal with the complications of the exact Coulomb Green's function/Coulomb wavefunctions.

With a view towards generalizing these results to QFT, we also consider the Heisenberg picture and the LSZ formulation of scattering. In quantum field theory, scattering amplitudes are commonly computed in the Heisenberg picture using the LSZ reduction formula, which relates S-matrix elements to time-ordered correlation functions \cite{Lehmann:1954rq}. For simplicity, we state the assumptions underlying the LSZ construction for a scalar Klein--Gordon field \(\phi\). The standard construction relies on two basic assumptions:
\begin{enumerate}
\item The interacting field \(\phi\) approaches the free in- and out-fields in the following sense. Let \(f\) be a solution of the free Klein-Gordon equation, and let \(\Sigma_t\) be a Cauchy surface with induced metric \(h\) and future-directed unit normal \(n^\mu\). The symplectically-smeared field is defined by
\be
\Omega(\phi(t),f(t))
\defn
\int_{\Sigma_t}d^3x\,\sqrt{h}\,n^\mu
\left(\phi\nabla_\mu f-f\nabla_\mu\phi\right) \,.
\ee
The incoming asymptotic quantum field \(\op \phi_\inn\) is a free quantum field which satisfies the limit 
\be
\lim_{t\to -\infty}
\braket{ \psi_1 | \Omega\big(\op\phi(t)- \op\phi_\inn(t),f(t)\big) | \psi_2 } = 0 \,.
\label{eq:lszassumption}
\ee
The limit is understood in a suitable weak sense, i.e., the matrix elements between states \(\ket{\psi_1}, \ket{\psi_2}\) in (a dense subspace of) some Hilbert space \(\Hilb\). An analogous condition defines the outgoing field \(\op\phi_\out\) by taking \(t\to +\infty\). These limits allow the asymptotic free fields, and hence the corresponding in and out particle operators, to be defined in terms of the interacting field.
\item The asymptotic in and out fields, \(\op\phi_\inn\) and \(\op\phi_\out\), are represented on the same Hilbert space \(\Hilb\). In particular, the in and out particle states can be compared within a common Hilbert space representation, and the S-matrix is defined as a unitary operator on \(\Hilb\) which maps from the in-fields to the out-fields.
\end{enumerate}

In quantum field theories with long-range interactions, both assumptions are problematic. In massive scalar QED, for example, the charged matter field does not have free asymptotic evolution because of its long-range electromagnetic interactions \cite{Campiglia:2015qka,Campiglia:2019wxe}. Moreover, although the massless field has free asymptotic dynamics, the long time behaviour of the massless field (related to the memory effect) leads to distinct asymptotic states that generically do not belong to the same standard Fock representation\footnote{Note that the obstruction to both assumptions has a common origin in the long-range nature of interactions \cite{Kulish:1970ut}.}, see \cite{Ashtekar:1987tt,Ashtekar:2018lor,Prabhu:2022zcr,Prabhu:2024zwl,Prabhu:2024lmg}. The non-relativistic Coulomb problem considered here isolates the first of these difficulties, since there are no propagating massless degrees of freedom. The Stone-von Neumann theorem ensures that we remain in a single Hilbert space which is unitarily equivalent to \(\Hilb = L^2(\mathbb R^3)\), and there is no obstruction from inequivalent Hilbert space representations. The difficulty instead lies in the asymptotic behaviour of the canonical observables as seen from the asymptotic classical trajectory \cref{eq:intro-x-falloff}.

To construct suitable asymptotic observables, we change from the phase space variables \((\vec x,\vec p)\) to new canonical variables \((\vec X,\vec P)\) using a time-dependent canonical transformation (see \cref{sec:heisenberg-LSZ} for details). The canonical transformation shifts the position variable by precisely the momentum-dependent term that gives rise to the logarithmic correction to the Coulomb trajectory. For the new variables $(\vec X, \vec P)$, we find the asymptotic behaviour as
\be
    \vec X(t) \sim \frac{\vec P_\out}{m} t + \vec X_\out + O(\ln t/t) \eqsp \vec P(t) \sim \vec P_\out + O(1/t) \,.
\ee
Thus, the evolution of the new variables asymptotes to a free solution and can therefore serve as the asymptotic observables in an LSZ-type construction. In \cref{sec:heisenberg-LSZ}, we show that this canonical transformation can be implemented by a unitary operator $\op{\mc W}(t)$, which is closely related to Dollard's modification (see \cref{eq:scW-defn}), so that the new Heisenberg operators \(\lb( \vec{\op X}(t), \vec{\op P}(t) \rb)\) are given by
\be
    \lb( \vec{\op X}(t), \vec{\op P}(t) \rb) &= \op{\mc W}(t) \lb( \vec{\op x}(t), \vec{\op p}(t) \rb) \op{\mc W}(t)^* \,.
\ee
The unitary evolution operator \(\op{\mc U}(t)\) for the new variables is
\be
    \op{\mc U}(t) \defn \op U(t) \op{\mc W}(t)^* \,,
\ee
where $\op U(t)$ is the ordinary Coulomb evolution operator for the original Heisenberg operators $(\op{\vec x}(t), \op{\vec p}(t))$. In \cref{sec:LSZ} we argue that the LSZ limit \cref{eq:lszassumption} assumption holds for new canonical variables, that is,
\be
    \lim_{t \to -\infty} \Omega\big( \vec{\op X}(t) - \vec{\op X}_\inn(t), \vec x_\free(t) \big) = 0 \,.
\ee
where $\vec x_{\free}$ is any free solution. We also show that the exact Coulomb wavefunctions $|\varphi_k^{\pm}\rangle $ are the (improper) eigenstates of the ``asymptotic'' in/out momenta operators $\op{\vec P}_{\rm out/in}$. With this definition of the asymptotic free operators, we can define the Heisenberg S-matrix \(\op S_{\rm H}\) as a map from the algebra of in-operators to the algebra of out-operators \cite{Morchio:2014zga}. In terms of the asymptotic position and momentum operators, we have
\be
    \lb( \vec{\op X}_\out, \vec{\op P}_\out \rb) = \op S_{\rm H}^* \lb(\vec{\op X}_\inn, \vec{\op P}_\inn \rb) \op S_{\rm H}
\ee
And the matrix elements of \(\op S_{\rm H}\) in the in (or out) Coulomb basis give the scattering amplitudes
\be
    \braket{\varphi_k^- | \op S_{\rm H} | \varphi_{k'}^-} &=  \braket{\varphi_k^+ | \op S_{\rm H} | \varphi_{k'}^+} = \braket{k | \op S | k'}
\ee
Thus the Heisenberg picture S-matrix amplitudes computed between the Coulomb in (or out) wavefunctions are the same as the Schrödinger picture amplitudes computed between free momentum eigenstates. Using this natural mapping, the Heisenberg picture S-matrix can be put in a similar form as \cref{eq:intosmatrix} but in the in/out variables, and our perturbative computation can be done directly in the Heisenberg picture.
 
We expect that the modified LSZ construction developed here can be straightforwardly generalized to the relativistic theory of a massive Klein--Gordon field in a static Coulomb potential. In QED with massive charged fields and dynamical photons, virtual photons running in loops of Feynman diagrams give rise to real divergences in addition to the imaginary Coulomb-type divergences. Moreover, the massless photons propagating to asymptotic null infinity generate a memory effect which forces the out-state to live in an inequivalent Hilbert space to the chosen in-state, which violates assumption (2) of the LSZ formalism, which also causes a divergence in the amplitudes. The structure of these inequivalent memory Hilbert spaces has been studied extensively \cite{Ashtekar:1987tt,Ashtekar:2018lor} along with their role in scattering theory, including conservation of asymptotic charges and soft theorems \cite{Prabhu:2022zcr,Prabhu:2024lmg,Prabhu:2024zwl}. We expect that the LSZ formalism can be generalized to include these asymptotic Hilbert spaces along with the construction of the asymptotic field operators along the lines of this paper; related perspectives on long-range dressing are discussed in \cite{Campiglia:2015qka,Campiglia:2019wxe,Choi:2019fuq,Choi:2024ajz,Choi:2024mac}. We are currently investigating these issues in the QFT context.

A complementary approach to defining the scattering operator is provided by the Aref'eva--Faddeev--Slavnov (AFS) path integral \cite{Arefeva:1974jv}, which is particularly natural from the perspective of asymptotic symmetries and boundary data, with the boundary terms on the asymptotic boundaries playing an essential role in the construction \cite{Fabbrichesi:1993kz,Kim:2023qbl,Jain:2023fxc,Upadhyay:2025ged,Isen:2026xoc}. The AFS path integral with the standard free boundary conditions generates the Dyson S-matrix, and its equivalence with the standard LSZ prescription was established in \cite{Kim:2023qbl}. In theories with long-range interactions, however, the fields appearing in the Lagrangian do not become free asymptotically. Nevertheless, as demonstrated in this paper, one can construct asymptotically free variables through a time-dependent canonical transformation. This suggests modifying the asymptotic boundary conditions of the AFS path integral or, alternatively, using the new canonical fields to define the path integral to construct an IR-finite S-matrix (see also \cite{Ammon:2026svb}). We hope to return to these questions in the future.

The rest of the paper is organized as follows. In \cref{sec:Coulomb-review}, we review the basics of the Coulomb scattering problem and the Dollard construction of the modified asymptotic dynamics. In \cref{sec:Born-series}, we develop the perturbative formulation of the Dollard scattering operator and demonstrate the cancellation of IR divergences through second Born order. In \cref{sec:heisenberg-LSZ}, we construct asymptotically free canonical variables and use them to formulate the scattering theory in the Heisenberg picture. We obtain the Heisenberg-picture S-matrix using several complementary methods and, finally, formulate the corresponding LSZ reduction procedure. In \cref{sec:conventions} we collect our conventions and useful distributional identities used throughout. In \cref{sec:Herbst} we review Herbst's analysis of the Coulomb S-matrix as a well-defined distribution on the Hilbert space, including its forward-direction contribution, and derive the perturbative expansion of this distributional amplitude in the coupling. In \cref{sec:Secondborn} we evaluate the regulated second Born integral using Mellin-Barnes techniques.

%%====================================================
\section{Coulomb scattering problem}
\label{sec:Coulomb-review}

In this section, we recall some well-known results about the Coulomb scattering problem. This will set up our conventions and also provide a reference point for the perturbative calculations that follow.

We consider a particle of mass $m$ scattering from a fixed, heavy Coulomb source located at \(\vec x = 0\). The total Hamiltonian is
\be
    H = H_0 + V \eqsp H_0 = \frac{ p^2}{2m} \eqsp V(\vec x) = \frac{\alpha}{\abs{\vec x}} \,.
\label{eq:coulomb-H}
\ee
We will take \(\alpha > 0\), i.e., a repulsive Coulomb potential, since we are interested in the scattering problem.

The exact Coulomb scattering eigenstate wavefunctions in position space\footnote{In momentum space the Coulomb wavefunctions \(\varphi_k^\pm(\vec p)\) are bi-distributions in \(\vec k\) and \(\vec p\) given explicitly in \cite{Guth-Mullin}.} are \cite{Landau:Lifschitz,Lippstreu:2023vvg}
\be
    \ket{\varphi_k^\pm} \equiv \varphi_k^\pm(\vec x) &= \frac{1}{(2\pi)^{\nfrac32}} \Gamma(1\mp i \gamma) e^{-\tfrac{\pi\gamma}{2}} e^{i \vec k \cdot \vec x} {}_1F_1(\pm i \gamma ; 1 ; \mp i k \abs{\vec x} - i \vec k \cdot \vec x ) \eqsp \gamma(k) = \frac{\alpha m}{k} \,.
\ee
We have put an additional factor of \(1/(2\pi)^{\nfrac32}\) so that for \(\alpha = 0\) these reduce to the usual momentum state in our conventions (see \cref{sec:conventions}) . The total energy eigenvalue of these states is
\be
    \op H \ket{\varphi_k^\pm} = \frac{k^2}{2m} \ket{\varphi_k^\pm} \,.
\label{eq:Coulomb-energy}
\ee
The two sets of scattering states are related by complex conjugation according to
\be
    \bar{\varphi_k^+}(\vec x) = \varphi_{-k}^-(\vec x) \eqsp \bar{\varphi_k^-}(\vec x) = \varphi_{-k}^+(\vec x) \,,
\ee
and are ``normalized'' according to
\be
    \braket{\varphi^+_k | \varphi^+_{k'}} = \braket{\varphi^-_k | \varphi^-_{k'}} &= \delta(\vec k - \vec k' ) \,.
\label{eq:Coulomb-orthonormal}
\ee
Either \(\{\ket{\varphi_k^+}\}\) or \(\{\ket{\varphi_k^-}\}\) is a complete set of \emph{improper} basis for the Hilbert space \(\Hilb \cong L^2(\bb R^3)\).

For a state \(\ket{\psi} \in \Hilb\) we can convert its components in the \(\{\ket{\varphi_k^-}\}\) basis to its components in the \(\{\ket{\varphi_k^+}\}\) basis as
\be
    \braket{\varphi_k^+ | \psi} = \int d^3k'~ \braket{\varphi_k^+ | \varphi_{k'}^-} \braket{\varphi_{k'}^- | \psi}
\ee
This inner product between the in and out Coulomb basis states defines the scattering amplitude \cite{Lippstreu:2023vvg,Landau:Lifschitz}
\be
    S(\vec k, \vec k') &= \braket{\varphi_k^+ | \varphi_{k'}^- } = \int d^3 x~ \bar{\varphi_k^+}(\vec x)\varphi_{k'}^-(\vec x) 
\label{eq:S-scattering-basis}
\ee

This expression can be explicitly computed. Let \(E_k = k^2/(2m)\) be the energy associated to the momentum \(\vec k\); note from \cref{eq:Coulomb-energy} this is the \emph{total energy} of the Coulomb scattering states, not just the kinetic energy. Further let \(\theta\) be the relative angle between \(\vec k\) and \(\vec k'\). Then, we have \cite{Lippstreu:2023vvg,Landau:Lifschitz}
\be
    S(\vec k, \vec k') &= -\frac{i \alpha}{4\pi k^2} \delta (E_k - E_{k'}) \frac{\Gamma (1+i\gamma)}{\Gamma(1-i\gamma)}  \frac{ e^{-i\gamma \ln \sin^2 \tfrac{\theta}{2}} }{\sin^2 \tfrac{\theta}{2} } \eqsp \theta \neq 0 \\[1.5ex]
    &= -\frac{i \alpha}{4\pi k^2} \delta (E_k - E_{k'}) \frac{\Gamma (1+i\gamma)}{\Gamma(1-i\gamma)}  \lb( \sin^2 \tfrac{\theta}{2} \rb)^{-1 - i \gamma } \eqsp \theta \neq 0
\label{eq:amplitude-naive}
\ee
where, we recall, \(\gamma = \gamma(k) = \gamma(k') = \alpha m / k\). \Cref{eq:amplitude-naive} differs from the Rutherford amplitude only by an overall phase. We emphasize that the restriction \(\theta \neq 0\) is essential in \cref{eq:amplitude-naive} since the expression is singular and not even integrable at \(\theta = 0\), the \emph{forward scattering}. We further note that the \(\alpha \to 0\) limit does not reproduce the free scattering amplitude, especially since forward scattering is explicitly excluded.

It has been often argued that the forward scattering must be excluded on physical grounds. However, this is unsatisfactory if the amplitude \cref{eq:amplitude-naive} is to arise from a genuine unitary operator on the Hilbert space \(\Hilb\). The primary reason for this issue is that \(S(\vec k, \vec k')\) is defined by the inner product of non-normalizable (improper) states (\cref{eq:S-scattering-basis}). This inner product should be understood as a distribution and \emph{not} an ordinary function. For a ``short range/short lived'' potential the distributional nature of the amplitude is concentrated entirely in the delta function of momenta at zeroth order in coupling, i.e., the forward direction singularity of the free scattering. However, the distributional nature of the Coulomb scattering amplitude is not quite so simple and has been analyzed in detail in \cite{Herbst:1974hi} by Herbst.\footnote{Taylor \cite{Taylor:1973da} has also considered the distributional convergence of the partial waves amplitude. However, the forward direction, \(\theta = 0\), was still excluded since the Coulomb amplitude itself was not considered as a distribution.} We recall this analysis in \cref{sec:Herbst} and also obtain a distributional perturbative expansion of the amplitude. See \cite{Lippstreu:2026xfw} for a recent discussion on the subtleties of forward scattering in theories with long-range interactions.

Next, we recall Dollard's construction of the S-matrix operator \(\op S\), whose matrix elements in the momentum basis reproduce the above amplitude.

%%--------------------------------------------------------
\subsection{Dollard's formulation of the S-matrix in the Schr\"odinger picture}
\label{sec:Dollard}

We first review the construction of the Dollard scattering operator. The essential point is that the usual scattering theory assumes that the interacting dynamics approaches free dynamics at large times. This assumption fails for long-range potentials such as the Coulomb potential. Dollard modified the asymptotic dynamics so that the long-range part of the
interaction is incorporated into the reference evolution, allowing the M\o ller operators to exist as strong limits \cite{Dollard:1964cok}.

We define the \emph{Dollard potential} --- see also Herbst \cite{Herbst:1974hi} (who uses \(m = 1\)) and Reed-Simon \cite{Reed-Simon3} (who use \(m = 1/2\))
\be
    V_{\rm D}(p, t) &\defn V(\vec p t/m)~ \theta(\abs{4 H_0t} - 1 ) = \frac{\gamma}{\abs{t}} ~ \theta(\abs{4 H_0t} - 1 ) \eqsp \text{where} \\
    \gamma = \gamma(p) &~= \frac{\alpha m}{p} \eqsp H_0 = \frac{p^2}{2m} \,.
\label{eq:V-Dollard}
\ee
Note that \(V_{\rm D}\) is the Coulomb potential evaluated on a free particle trajectory \(\vec x = \vec p t/m\) and the step function avoids introducing a spurious singularity at \(t=0\). We elaborate on the choice of time scale \(4H_0\) in \cref{rem:Dollard-scale}.  We also introduce the \emph{Dollard Hamiltonian} \(H_{\rm D}(t)\) given by
\be
    H_{\rm D}(t) \defn H_0 + V_{\rm D} \,,
\ee
and the \emph{Dollard phase}
\be
   \Phi(p,t) \defn \int_0^t dt'~ V_{\rm D}(p,t') = \gamma \sgn(t) \ln \abs{4H_0 t}~ \theta(\abs{4 H_0t} - 1 ) \,.
\label{eq:Phi-Dollard}
\ee\\
In the quantum scattering problem define the unitaries
\be
    \op U(t) \defn e^{-i\op H t} \eqsp \op U_0(t) \defn e^{-i \op H_0 t} \eqsp \op{ W}(t) \defn e^{-i \Phi(\op p,t)} \,,\\
    \op U_{\rm D}(t) \defn \exp \lb( -i \int_0^t dt'~ \op H_{\rm D}(t') \rb) = \op W(t) \op U_0(t) \,.
\label{eq:evol-ops}
\ee
Dollard \cite{Dollard:1964cok} shows that, for every state \(\ket{\psi} \in \Hilb\) there exist states \(\ket{\psi_\pm} \in \Hilb\) such that 
\be
    \lim_{t \to \pm\infty} \norm{ \op U(t)\ket{\psi} - \op U_{\rm D}(t)\ket{\psi_\pm} } = 0 \,,
\label{eq:Dollard-limit}
\ee
i.e., asymptotically, the evolution of \(\ket{\psi}\) matches the Dollard evolution of \(\ket{\psi_\pm}\), in the strong sense. This implies the existence of the M\o ller operators which are\footnote{For ``short range/short lived'' potentials \(\op U_{\rm D}(t)\) can be replaced by \(\op U_0(t)\) in \cref{eq:Dollard-limit,eq:Moller-defn}. However for the Coulomb potential the weak limit vanishes, \(\wlim_{t \to \pm \infty} \op U(t)^* \op U_0(t) = 0 \), and thus the strong limit does not exist \cite{Reed-Simon3}.}
\be
    \op\Omega_\pm \defn \slim_{t \to \pm\infty} \op U(t)^* \op U_{\rm D}(t) \eqsp \ket{\psi} = \op\Omega_\pm \ket{\psi_\pm}
\label{eq:Moller-defn}
\ee
where ``\(\slim\)'' indicates the strong limit of operators. Since the M\o ller operators are strong limits of unitaries they are also unitary operators\footnote{If we take an attractive Coulomb potential, \(\alpha < 0\), we must allow for bound states. Then the M\o ller operators \(\op\Omega_\pm\) are \emph{partial isometries} or \emph{partial unitaries} on \(\Hilb\) whose range is the positive total energy subspace \(\Hilb_+ \subset \Hilb\). Then, \(\op\Omega_\pm^*\) are defined with the domain \(\Hilb_+\), and \cref{eq:Moller-unitary} is changed to \(\op\Omega_\pm^* \op\Omega_\pm = \op P_+\) and \(\op\Omega_\pm \op\Omega_\pm^* = \1\) where \(\op P_+\) is the projection onto the positive energy subspace.} on \(\Hilb\), i.e.,
\be
    \op\Omega_\pm^* \op\Omega_\pm = \op\Omega_\pm \op\Omega_\pm^* = \1
\label{eq:Moller-unitary}
\ee
and further satisfy the intertwining relation \cite{Dollard-review, Morchio:2014zga}
\be
    \op U(t) \op\Omega_\pm = \op\Omega_\pm \op U_0(t) \,.
\label{eq:Moller-intertwine}
\ee

Thm.~2 of \cite{Dollard:1964cok} also shows that for \(\ket{\psi}\) in a dense subspace of \(\Hilb\) with \(\psi(k) = \braket{k|\psi}\) the Fourier space wavefunction, the M\o ller operators act as
\be
    (\op\Omega_\pm \psi )(x) &= \int d^3k~ \varphi_k^\pm(x) \psi(k) \implies
    \op\Omega_\pm\ket{\psi} &= \int d^3k~ \ket{\varphi_k^\pm} \braket{k|\psi} \,,
\label{eq:Moller-on-state}
\ee
from which we get the basis expressions
\be
    \op\Omega_\pm = \int d^3k~ \Dop{\varphi_k^\pm}{k} \eqsp \op\Omega_\pm^* = \int d^3k~ \Dop{k}{\varphi_k^\pm} \,,
\label{eq:Moller-basis}
\ee
and
\be
    \op\Omega_\pm \ket{k} = \ket{\varphi_k^\pm} \eqsp \op\Omega_\pm^* \ket{\varphi_k^\pm} = \ket{k} \,,
\label{eq:Moller-improper}
\ee
in the sense of improper states.

The Schr\"odinger-picture S-matrix \(\op S\) is defined as the map from in-states \(\ket{\psi_-}\) to out-states \(\ket{\psi_+}\)
\be
    \ket{\psi_+} = \op S \ket{\psi_-} \,,
\ee
and using \cref{eq:Moller-defn,eq:Moller-basis} we have
\be
    \op S \defn \op\Omega_+^* \op\Omega_- =  \int d^3k d^3k'~ \ket{k} \braket{\varphi_k^+ | \varphi_{k'}^-} \bra{k'} \,,
\label{eq:S-defn}
\ee
which also shows that \(\op S\) is a unitary. The scattering amplitude are then
\be
    S(\vec k, \vec k') = \braket{k | \op S | k'} = \braket{\varphi_k^+ | \varphi_{k'}^-} \,,
\ee
which matches the definition \cref{eq:S-scattering-basis}.

\begin{remark}[Dollard time scale]
    As emphasized in \cite{Dollard:1964cok}, the choice of time scale \(4 H_0\) in \cref{eq:V-Dollard}, does not affect any of the convergence arguments at large time. The particular choice is made so that \cref{eq:Moller-on-state} holds, i.e., action of the M\o ller operator reproduces the components of the state in the Coulomb scattering basis. Suppose we use a different time scale \(T\), which can depend on the magnitude of the momentum, in the Dollard potential \(V_{\rm D}\) then we get
\be
    \tilde V_{\rm D}(p,t) = \frac{\gamma(p)}{|t|}\theta(|t/T|-1) = \frac{1}{4H_0 T} V_{\rm D} \lb(p,\frac{t}{4H_0 T}\rb) \eqsp
    \tilde\Phi(p,t) = \Phi\lb(p,\frac{t}{4H_0T}\rb) \, .
\ee
Thus, at large \(|t|\) we have
\be
    \tilde{\op W}(t) \sim \op W(t) e^{+i \sgn(t) \beta(\op p, T)} \eqsp \tilde{\op\Omega}_\pm = \op \Omega_\pm e^{\pm i \beta(\op p, T)} \eqsp \tilde{\op{S}} = e^{- i \beta(\op p, T)} \op S e^{- i \beta(\op p, T)} 
\ee
where \(\beta(p, T) = \gamma(p) \ln(4H_0 T)\). Note that, since \(\op S\) conserves energy, the ambiguity in the S-matrix is an overall phase which may depend on the magnitude of the momentum. From \cref{eq:Moller-improper}, this phase can be seen as an ambiguity in the relative phase between the in and out Coulomb scattering states, e.g., \(\ket{\tilde \varphi_k^\pm} = e^{\pm i \beta(k, T)} \ket{\varphi_k^\pm}\). We note that both \(\ket{\varphi_k^\pm}\) and \(\ket{\tilde\varphi_k^\pm}\) are eigenstates of the same Coulomb Hamiltonian (\cref{eq:Coulomb-energy}), with the same energy eigenvalue. The ``normalization'' \cref{eq:Coulomb-orthonormal} only fixes each set of eigenbasis up to a momentum dependent phase, and there is no way to fix the relative phase between the in and out basis purely from the scattering problem. Lippstreu \cite{Lippstreu:2025jit} has criticized Dollard's construction due to the arbitrariness of this time scale. As we have argued above the same time scale appears in the exact Coulomb scattering eigenstates as a relative phase ambiguity. Instead the analytic properties described in \cite{Lippstreu:2025jit} can be used as additional conditions to fix the choice of this time scale.
\label{rem:Dollard-scale}
\end{remark}

%%========================================================
\section{Perturbative computation of the Coulomb amplitude}
\label{sec:Born-series}

In this section we show that the Coulomb S-matrix amplitude can be computed perturbatively in the coupling constant \(\alpha\) when Dollard's modification to the asymptotic dynamics is taken into account. We will work with the Coulomb potential without directly using the Coulomb scattering wavefunctions or the Coulomb Green's function.

In standard scattering theory with a ``short range/short lived'' potential \(V\), the S-matrix is given by
\be
\op S = \Tord{ \exp\lb( - i \int_{-\infty}^{+\infty} dt~ \op V_{\rm I}(t) \rb)}
\label{eq:S-usual}
\ee
where $\op V_{\rm I}(t) = \op U_0(t)^* \op V \op U_0(t)$ is the ``interaction picture'' potential. The scattering amplitude is obtained as matrix elements of \(\op S\) between states in a complete basis of the Hilbert space. The amplitude can be computed perturbatively in the interaction strength, or coupling constant, which gives the Dyson series (or the Born series for static potentials). For the Coulomb potential, however, the integral of the potential in \cref{eq:S-usual} diverges; this divergence in the S-matrix causes the integrals in the na\"ive Born-Dyson series to also diverge.

We now give an explicit computation of the Coulomb S-matrix taking into account Dollard's modification following \cref{sec:Dollard}. For our computations we will regularize the Coulomb potential as follows
\be
    V_\epsilon(\vec x) = \frac{\alpha \lambda^{2\epsilon}}{\abs{\vec x}^{1+2\epsilon}} \eqsp V_\epsilon(\vec p) = \frac{1}{(2\pi)^\half} \frac{2 \alpha \lambda^{2\epsilon}}{p^{2-2\epsilon}} \Gamma(1-2\epsilon) \cos (\epsilon\pi) \\
\label{eq:FourierV}
\ee
where $\epsilon>0$ and \(\lambda\) is an arbitrary length scale introduced so that \(V_\epsilon\) has the correct dimensions of a potential. We note that for \(\epsilon > 0\), the scattering off the potential \(V_\epsilon\) does have a well-defined S-matrix, and Born series, without any need for Dollard's modification \cite{Dollard-review}. But since we are interested precisely in the \(\epsilon \to 0\) case, we similarly regularize the Dollard potential (\cref{eq:V-Dollard}) as
\be
    V_{\rm D,\epsilon}(\vec p, t) = V_\epsilon(\vec p t/m) \theta(|4 H_0 t| - 1) \eqsp V_{\rm D,\epsilon}(p,\omega = 0) = \frac{\gamma}{(2\pi)^\half} \frac{(2p\lambda)^{2\epsilon}}{\epsilon} \,,
\label{eq:FourierVD}
\ee
where the second expression is the Fourier transform at zero frequency \(\omega = 0\) --- we will only need the zero-frequency value of \(V_{\rm D,\epsilon}\) for our computation.

We first compute the S-matrix with the regulator \(\epsilon > 0\), including Dollard's modification. We show that taking \(\epsilon \to 0^+\) in the end\footnote{We have not attempted to show that the \(\epsilon \to 0^+\) limit commutes with the strong limits of operators as \(t \to \pm \infty\). We expect this to be true since our procedure reproduces the expected perturbative amplitude from the exact one.} gives a well-defined S-matrix, which is also independent of the arbitrary length scale \(\lambda\) introduced in \cref{eq:FourierV}. Perturbatively, this computation also reproduces the distributional series expansion of the exact Coulomb amplitude in \cref{sec:Herbst}.

Using \cref{eq:evol-ops,eq:Moller-defn}, we have
\be
    \op\Omega_{-,\epsilon} &= \slim_{t \to -\infty} \op U_\epsilon(t)^* \op U_{\rm D, \epsilon}(t) = \slim_{t \to -\infty} \op U_\epsilon(t)^* \op U_0(t) \op W_\epsilon(t) =  \op U_{\rm I, \epsilon}(-\infty)^* \op W_\epsilon(-\infty) \\
    \op\Omega_{+,\epsilon}^* &= \slim_{t \to +\infty} \op U_{\rm D, \epsilon}(t)^* \op U_\epsilon(t) = \slim_{t \to +\infty} \op W_\epsilon(t)^* \op U_0(t)^* \op U_\epsilon(t) =  \op W_\epsilon(+\infty)^* \op U_{\rm I, \epsilon}(+\infty) \,.
\ee
Here \(\op U_{\rm I,\epsilon}\defn \op U_0(t)^* \op U_\epsilon(t)\) is the interaction-picture evolution operator for the regulated potential. It is possible to absorb the Dollard correction into an effective evolution operator and then apply standard perturbation theory to this effective interaction. However, this effective interaction potential is quite complicated and also obscures the explicit factorization of the divergent Coulomb phase. 

Therefore, the S-matrix, defined in \cref{eq:S-defn} can be written as
\be
    \op S_\epsilon &= \op\Omega_{+,\epsilon}^* \op\Omega_{-\epsilon} = \op W_\epsilon(+\infty)^* \op U_{\rm I, \epsilon}(+\infty) \op U_{\rm I, \epsilon}(-\infty)^* \op W_\epsilon(-\infty) \\
    &= \exp \lb( + i \int_0^{+\infty} dt~ \op V_{\rm D, \epsilon}(t) \rb) \Tord{ \exp\lb( - i \int_{-\infty}^{+\infty} dt~ \op V_{\rm I, \epsilon}(t) \rb)} \exp \lb( + i \int_{-\infty}^0 dt~ \op V_{\rm D, \epsilon}(t) \rb)
\ee
In frequency space we have (from \cref{eq:FT-half})
\be
    \int_0^{+\infty} dt~ \op V_{\rm D, \epsilon}(t) & = \frac{1}{(2\pi)^\half} \int_{-\infty}^{+\infty} d\omega~ \frac{-i \op V_{\rm D,\epsilon}(\omega)}{\omega - i0^+} \eqsp \int_{-\infty}^0 dt~ \op V_{\rm D, \epsilon}(t) &= \frac{1}{(2\pi)^\half} \int_{-\infty}^{+\infty} d\omega~ \frac{+i \op V_{\rm D,\epsilon}(\omega)}{\omega+ i0^+}
\ee
Then, since the Dollard potential is diagonal in momentum, we compute the amplitude as matrix elements between (improper) momentum eigenstates to get
\be
    \braket{p_2 | \op S_\epsilon | p_1}
    &= \exp \lb(\frac{1}{(2\pi)^\half} \int_{-\infty}^{+\infty} d\omega~ \frac{V_{\rm D,\epsilon}(p_2,\omega)}{\omega - i0^+} \rb) \braket{p_2 | \Tord{ \exp\lb( - i \int_{-\infty}^{+\infty} dt~ \op V_{\rm I, \epsilon}(t) \rb)} | p_1} \\
    &\qquad \times \exp \lb( - \frac{1}{(2\pi)^\half} \int_{-\infty}^{+\infty} d\omega~ \frac{V_{\rm D,\epsilon}(p_1,\omega)}{\omega+ i0^+} \rb)
\ee
Now the regulated potential \(V_\epsilon\) is static and so conserves energy, and the Dollard potential is a function only of the magnitude of momentum. So we can combine the two exponentials of the Dollard potential, using \cref{eq:SP}, to get
\be
    \int_{-\infty}^{+\infty} d\omega~ \frac{V_{\rm D,\epsilon}(p,\omega)}{\omega - i0^+} - \int_{-\infty}^{+\infty} d\omega~ \frac{V_{\rm D,\epsilon}(p,\omega)}{\omega+ i0^+}
    = 2 \pi i V_{\rm D,\epsilon}(p,0) \,,
\ee
where \(p = p_1 = p_2\). Consequently
\be
    \braket{p_2 | \op S_\epsilon | p_1}
    &= \exp \lb( i (2 \pi)^\half V_{\rm D,\epsilon}(p,0) \rb) \braket{p_2 | \Tord{ \exp\lb( - i \int_{-\infty}^{+\infty} dt~ \op V_{\rm I, \epsilon}(t) \rb)} | p_1}
\label{eq:perturbativeSmatrix}
\ee
Thus, the Coulomb S-matrix can be expressed as the Dyson S-matrix multiplied by a phase. Note that both factors are individually divergent as \(\epsilon\to 0^+\), so the regulator must be removed only after combining them. Further, the phase correction from the Dollard potential is at zero frequency, i.e., the issue of Coulomb divergences is really an infinite-time effect, as is expected from Dollard's analysis\footnote{We also expect this to be true in relativistic quantum field theories with massless fields where in addition to divergences from virtual particles running in loop, we also have the soft divergence due to real emissions, see \cite{Prabhu:2022zcr,Prabhu:2024lmg,Prabhu:2024zwl}.}. We also emphasize that the advantage of putting the S-matrix in this form is that it avoids explicitly using the exact Coulomb Green's functions or Coulomb scattering wavefunctions while keeping the cancellation of the divergences explicit. 

Before proceeding to the perturbative computations, we argue that even non-perturbatively \cref{eq:perturbativeSmatrix} reproduces the exact Coulomb S-matrix in the limit \(\epsilon \to 0^+\). For our purposes\footnote{The same result can be obtained using an eikonal analysis; see \cite{Kabat:1992tb,Nekrasov:2023kxu}.} it will be useful to use the analysis of Taylor \cite{Taylor:1973da}, see also \cite{Dalitz:1951ah,Weinberg:1965nx}. Using the partial wave basis Taylor shows that the exact Coulomb amplitude is related to the amplitude of the regulated potential by a phase (with \(r = \abs{\vec x}\), also note that Taylor uses the convention \(m = p = 1\))
\be
    \braket{p_2 | \Tord{ \exp\lb( - i \int_{-\infty}^{+\infty} dt~ \op V_{\rm I, \epsilon}(t) \rb)} | p_1} ~\underset{\epsilon \to 0^+}{\sim}~ \exp\lb[- 2 i \frac{m}{p} \int_{\tfrac{1}{2p}}^\infty  dr~ V_\epsilon(r) \rb] \braket{p_2 | \op S | p_1}
\label{eq:Taylor-result}
\ee
where the limit is understood in the sense of distributions. We wish to use this result for our regulated potential \(V_\epsilon\) (\cref{eq:FourierV}). However, Taylor's analysis assumes that the regulated (or screened) Coulomb potential falls-off as \(O(r^{-2 - \epsilon})\) as \(r \to \infty\), which is not true for our regulated potential. However, we do expect that the estimates used by Taylor \cite{Taylor:1973da} can also be proven for our \(V_\epsilon\), though we have not shown this. Proceeding by assuming Taylor's result \cref{eq:Taylor-result} holds, we get (using \cref{eq:FourierV})
\be
    \braket{p_2 | \Tord{ \exp\lb( - i \int_{-\infty}^{+\infty} dt~ \op V_{\rm I, \epsilon}(t) \rb)} | p_1} ~\underset{\epsilon \to 0^+}{\sim}~ \exp\lb[- i \frac{\gamma}{\epsilon} (2p \lambda)^{2\epsilon} \rb] \braket{p_2 | \op S | p_1} \,.
\ee
Comparison with \cref{eq:perturbativeSmatrix} and using \cref{eq:FourierVD} we get
\be
    \lim_{\epsilon \to 0^+} \braket{p_2 | \op S_\epsilon | p_1} = \braket{p_2 | \op S | p_1} \,.
\ee
That is, the regulated Coulomb S-matrix, with the Dollard modification, limits to the exact Coulomb S-matrix.

Next we show that one can also perturbatively compute the Coulomb amplitude using \cref{eq:perturbativeSmatrix}. We will explicitly compute the first and second order in the coupling constant \(\alpha\), and show that it admits a well-defined distributional limit as \(\epsilon \to 0^+\). As emphasized before this computation does not use pre-existing knowledge of the exact Coulomb scattering basis, or Coulomb Green's functions. Nevertheless, it reproduces the distributional expansion of the exact Coulomb amplitude computed in \cref{sec:Herbst}. We also expect a similar perturbative computation to be useful for resolving the IR divergences in QFT, where the exact non-perturbative amplitude is, generally, not known.

 Since both $ V_\epsilon$ and $ V_{\rm D,\epsilon}$ are $O(\alpha)$, we expand both the Dollard exponential and the Dyson exponential to second order
\be
e^{i(2\pi)^\half  V_{\rm D,\epsilon}({p}_2,0)} =
1 + i(2\pi)^\half V_{\rm D,\epsilon}({p}_2,0) -\pi V_{\rm D, \epsilon}^2(p_2,0) + O(\alpha^3).
\label{eq:VD-expansion}
\ee
and
\be
    \braket{p_2 | \Tord{ \exp\lb( - i \int_{-\infty}^{+\infty} dt~ \op V_{\rm I,\epsilon}(t) \rb)} | p_1}
&=
\delta(\vec p_2-\vec p_1)
-
i\int_{-\infty}^{\infty}dt\,
\braket{ {p}_2|\op V_{\rm I, \epsilon}(t)|{p}_1 } \\
&\quad -
\int_{-\infty}^{\infty}dt_2
\int_{-\infty}^{t_2}dt_1\,
\braket{ {p}_2| \op V_{\rm I, \epsilon}(t_2) \, \op V_{ \rm I, \epsilon}(t_1)| p_1 }
+
O(\alpha^3).
\label{eq:Dysonexpansion}\ee
From these it is clear that the zeroth-order amplitude is simply the delta function in momenta, as expected for free scattering. We compute the next two orders in the following sections. We expect that similar computations can be done to higher orders in \(\alpha\).

%===================================================================================================
\subsection{First-order amplitude}
\label{sec:first-order}

The first-order amplitude is given by keeping only the first-order terms in \cref{eq:VD-expansion,eq:Dysonexpansion} gives
\be
\braket{ {p}_2|\op S_\epsilon^{(1)}|{p}_1 }
&=
-i\int_{-\infty}^{\infty}dt\,
\braket{ {p}_2|\op V_{\rm I, \epsilon}(t)|{p}_1 }
+i(2\pi)^{1/2}
 V_{D,\epsilon}({p}_2,0)
\delta(\vec {p}_2-\vec {p}_1).
\ee
Using 
\be
\op V_{\rm I, \epsilon}(t)=\op U_0(t)^* \op V_\epsilon \, \op U_0(t)
\ee
and the spatial Fourier transform of the potential,
\be
\braket{ {p}_2|\op V_\epsilon|{p}_1 }
=
\frac{1}{(2\pi)^{\nfrac32}} V_\epsilon(\vec {p}_2-\vec {p}_1),
\ee
we obtain
\be
\int_{-\infty}^{\infty}dt\,
\braket{ {p}_2|\op V_{\rm I, \epsilon}(t)|{p}_1 }
&=
\int_{-\infty}^{\infty}dt\,
e^{+i(E_2-E_1)t}
 V_\epsilon(\vec {p}_2-\vec {p}_1)
=
\frac{\delta(E_2-E_1)}{(2\pi)^\half}
 V_\epsilon(\vec{p}_2-\vec{p}_1) \,.
\ee
where \(E_i = p_i^2/(2m)\). So at the first order we get, 
\be
    \braket{p_2 | \op S_\epsilon^{(1)} | p_1} &= - i  \lb[(2\pi)^{-\half} V_\epsilon(\vec{p}_2-\vec{p}_1)
\delta(E_2 - E_1) - (2\pi)^\half V_{D,\epsilon}({p}_2,0) \delta(\vec {p}_2-\vec {p}_1) \rb] \\
    & =  -i \lb[ \frac{4 m \alpha \lambda^{2\epsilon}}{\abs{\vec p_2 - \vec p_1}^{2-2\epsilon}} \Gamma(1-2\epsilon) \cos\left({\epsilon\pi}\right) \frac{\delta(p_2^2 - p_1^2)}{2\pi} - \frac{\gamma (2 p \lambda)^{2\epsilon}}{\epsilon} \delta(\vec p_2 - \vec p_1) \rb] \\
    & =  -i \gamma (2 p\lambda)^{2\epsilon} \lb[ \Gamma(1-2\epsilon) \cos\left({\epsilon\pi}\right) \frac{\delta(p_2^2 - p_1^2)}{2\pi p \lb( \sin^2\tfrac{\theta}{2} \rb)^{1-\epsilon} } - \frac{1}{\epsilon} \delta(\vec p_2 - \vec p_1) \rb]
\ee
where in the second line we used \cref{eq:FourierV,eq:FourierVD}. Expanding in small \(\epsilon\)
\be
    \braket{p_2 | \op S_\epsilon^{(1)} | p_1} 
    & =  - i \gamma \lb[ (1 + 2\epsilon \gamma_{\rm E}) \frac{\delta(p_2^2 - p_1^2) }{ 2\pi p \lb(\sin^2 \tfrac{\theta}{2} \rb)^{1-\epsilon} } - \frac{\delta(\vec p_2 - \vec p_1)}{\epsilon} \rb] + O(\epsilon)
\ee
We \emph{cannot} take the limit \(\epsilon \to 0^+\) of the individual terms because while the second term diverges, even the first term does not converge to a distribution. But we can combine them as
\be
    \braket{p_2 | \op S_\epsilon^{(1)} | p_1} 
    & =  - i \gamma \bigg[ ( 1 + 2\epsilon \gamma_{\rm E} ) \lb( \frac{\delta(p_2^2 - p_1^2) }{ 2\pi p \lb(\sin^2 \tfrac{\theta}{2} \rb)^{1-\epsilon} } - \frac{\delta(\vec p_2 - \vec p_1)}{\epsilon} \rb) + 2 \gamma_{\rm E} \delta(\vec p_2 - \vec p_1) \bigg] + O(\epsilon) \,.
\ee
Now we can safely take the limit \(\epsilon \to 0^+\) to get the distribution (see \cref{eq:Pf-x})
\be
    \braket{p_2 | \op S^{(1)} | p_1} 
     & =  - i \gamma \lb[ 2 \gamma_{\rm E} \delta (\vec p_2 - \vec p_1) + \frac{\delta(p_2^2 - p_1^2)}{ 2\pi p} \Pf \lb(\sin^2\tfrac{\theta}{2}\rb)^{-1} \rb] \,.
\label{eq:firstorderamplitude}\ee
Here \(\Pf\) denotes the Hadamard finite part, and the first-order amplitude exactly matches the distributional Coulomb amplitude at the same order, \cref{eq:S1-dist}. Note that away from the forward direction it reproduces the Rutherford amplitude obtained from the usual Born series for the Coulomb potential. Thus, at first-order the Dollard modification serves to extend the na\"ive Born amplitude to the forward direction as a well-defined distribution.
%%-------------------------------------------------
\subsection{Second-order amplitude}
\label{sec:second-order}

For the second-order amplitude, from \cref{eq:VD-expansion,eq:Dysonexpansion}, we get three terms
\be
    \braket{p_2|\op S_\epsilon^{(2)}|p_1}
    &=-\pi V_{\rm D,\epsilon}^2(p_2,0)
\delta(\vec p_2-\vec p_1) + (2\pi)^\half V_{\rm D,\epsilon}(p_2,0)
\int_{-\infty}^{+\infty}dt\,
\braket{ p_2|\op V_{\rm I,\epsilon}(t)|p_1 } \\
    &\qquad-\int_{-\infty}^{+\infty}dt_2
\int_{-\infty}^{t_2}dt_1~
\braket{ p_2|
\op V_{\rm I,\epsilon}(t_2)
\op V_{\rm I,\epsilon}(t_1)
|p_1 } \,.
\label{eq:nettsecondorder}
\ee
Using the first-order matrix element, the mixed contribution can be written as
\be
    \braket{ p_2|\op S_{\epsilon,\rm VD}|p_1 } =
V_{\rm D,\epsilon}(p_2,0)\,
V_\epsilon(\vec p_2-\vec p_1)\,
\delta(E_2-E_1).
\ee
The pure Born-Dyson contribution is
\be
    \braket{ p_2|\op S_{\epsilon,\rm VV}^{(2)}|p_1 }
\defn{}&
-\int_{-\infty}^{+\infty}dt_2
\int_{-\infty}^{t_2}dt_1\,
    \braket{ p_2|
\op V_{\rm I,\epsilon}(t_2)
\op V_{\rm I,\epsilon}(t_1)
|p_1 } \,.
\label{eq:S2VV-matrix}
\ee
Inserting a resolution of the identity in the momentum basis and using the interaction-picture matrix elements gives
\be
    \braket{ p_2|\op S_{\epsilon,\rm VV}^{(2)}|p_1 }
={}&-\frac{1}{(2\pi)^3}
\int d^3\ell
\int_{-\infty}^{+\infty}dt_2
\int_{-\infty}^{t_2}dt_1\,
e^{i(E_2-\omega)t_2}
e^{i(\omega-E_1)t_1} V_\epsilon(\vec p_2-\vec\ell)\,
V_\epsilon(\vec\ell-\vec p_1) \\
    ={}&i\frac{2m}{(2\pi)^2}\,
\delta(E_2-E_1)
\int d^3\ell\,
\frac{
V_\epsilon(\vec p_2-\vec\ell)\,
V_\epsilon(\vec\ell-\vec p_1)
}{
\ell^2-p_2^2-i0^+
}
\ee
where \(\omega=\ell^2/(2m)\) and we have used \cref{eq:FT-step}. We compute this integral in \cref{sec:Secondborn}; the final answer is given by
\be
    \braket{ p_2 |\op S^{(2)}_{\epsilon, \rm VV} | p_1 } = - \gamma^2 \Gamma^2(1-2\epsilon)\cos^2(\epsilon\pi)  \frac{(2p\lambda)^{4\epsilon}}{\epsilon}  \frac{\delta(p_2^2 - p_1^2) }{ 2 \pi p} u^{-1 +2\epsilon} \lb[ 1 - \epsilon^2 R(u) + O(\epsilon^3)\rb]
\label{eq:VVcontribution}
\ee
where $p = \abs{\vec p_2} = \abs{\vec p_1}$, \(u=\sin^2\tfrac{\theta}{2}\) and
\be
    R(u)
    =
    \operatorname{Li}_2(u)
    +2\pi i\ln(1+ u^\half)
    +\ln u\,\ln(1-u) \,.
\label{eq:R-defn}
\ee
Here, we note that this Born-Dyson contribution diverges as \(\epsilon \to 0^+\), reflecting the well-known divergence of the second Born term in Coulomb scattering \cite{Dalitz:1951ah}. We now show that the other two terms coming from the Dollard modification cancel this divergence, giving a well-defined distribution at second order.

Summing all three contributions in \cref{eq:nettsecondorder}, using \cref{eq:FourierV,eq:FourierVD,eq:VVcontribution}, the total second-order amplitude is
\be
\braket{ p_2|\op S_{\epsilon}^{(2)}| p_1 } &= \gamma^2 (2p \lambda)^{4\epsilon} \left[-\frac{\delta(\vec p_2 - \vec p_1)}{2\epsilon^2} +\frac{\delta(p_2^2-p_1^2)}{2\pi p } \frac{\Gamma(1-2\epsilon) \cos(\epsilon \pi)}{\epsilon\, u^{1-\epsilon}} \right. \\
&\qquad \left.- \frac{\delta(p_2^2-p_1^2)}{2\pi p} \frac{\Gamma^2(1-2\epsilon) \cos^2(\epsilon \pi)}{\epsilon\,u^{1-2\epsilon}} (1-R(u) \epsilon^2 + O(\epsilon^3)) \right]
\label{eq:secondorder}
\ee
Evidently, the limit \(\epsilon \to 0^+\) does not exist as an ordinary function, but the limit \emph{does} exist in the sense of distributions. We use the distributional series expansion (\cref{eq:series}, see also \cite{Gelfand1})
\be
    \frac{\delta(p_2^2-p_1^2)} {2\pi p} \frac{1}{u^{1-\epsilon}}
&= \frac{\delta(\vec p_2-\vec p_1)}{\epsilon}
+ \frac{\delta(p_2^2-p_1^2)}{2\pi p}
\Pf\frac{1}{u} +
\epsilon\frac{\delta(p_2^2-p_1^2)}{2\pi p}
\Pf \frac{\ln u}{u} + O(\epsilon^2) \,,
\label{eq:pf-epsilon}
\ee
and a similar series with \(\epsilon \to 2\epsilon\). The expansions of Gamma functions are
\be
\Gamma(1-2\epsilon)\cos(\pi\epsilon)
&=1+2\gamma_{\rm E}\epsilon
+\left(2\gamma_{\rm E}^2-\frac{\pi^2}{6}\right)\epsilon^2
+O(\epsilon^3) \,, \\
\Gamma^2(1-2\epsilon)\cos^2(\pi\epsilon)
\left[1-\epsilon^2R(u)+O(\epsilon^3)\right]
&=
1+4\gamma_{\rm E}\epsilon
+\left(
8\gamma_{\rm E}^2-\frac{\pi^2}{3}-R(u)
\right)\epsilon^2
+O(\epsilon^3).
\label{eq:gamma-two-R-expansion}
\ee
where $\gamma_E$ is the Euler-Mascheroni constant. Before taking the distributional limit, we examine the behaviour
of $R(u)$. From \cref{eq:R-defn}, it follows that near \(u=0\) we have
\be
    \frac{R(u)}{u} = 2\pi i\,u^{-\half} - \ln u + O(1).
\ee
which is locally integrable near $u=0$. This implies
\be
\frac{\delta(p_2^2-p_1^2)}{2\pi p} \frac{R(u)}{u^{1-2\epsilon}} \to \frac{\delta(p_2^2-p_1^2)}{2\pi p} \frac{R(u)}{u} \,,
\ee
and in particular, it does not contribute a pole or delta-function. Now, substituting \cref{eq:pf-epsilon} into \cref{eq:secondorder} we see that the distributional series contains terms with $\delta(\vec p_2-\vec p_1)$, \(\Pf \tfrac{1}{u}\) and \(\Pf\tfrac{\ln u}{u}\) which we compute next. The $\delta(\vec p_2-\vec p_1)$ part is
\be
&\qquad \lim_{\epsilon \to 0^+} \frac{\gamma^2(2p\lambda)^{4\epsilon}}{\epsilon^2}
\bigg[
-\tfrac12
+\Gamma(1-2\epsilon)\cos(\pi\epsilon) \\
&\qquad\qquad -\tfrac12\Gamma^2(1-2\epsilon)\cos^2(\pi\epsilon)
(1+O(\epsilon^3))\bigg] \delta(\vec p_2-\vec p_1) \,, \\
    & = -2\gamma^2\gamma_{\rm E}^2 \delta(\vec p_2-\vec p_1) \,.
\label{eq:diagonal-finite}
\ee
Here we have used \cref{eq:gamma-two-R-expansion} and that $R(0)=0$, since the delta distribution is supported only at the forward point $u=0$.

The $\Pf\frac{1}{u}$ part is
\be
    &\qquad \lim_{\epsilon \to 0^+} \frac{\gamma^2(2p\lambda)^{4\epsilon}}{\epsilon}
\bigg[ \Gamma(1-2\epsilon)\cos(\pi\epsilon) \\
    &\qquad\qquad -\Gamma^2(1-2\epsilon)\cos^2(\pi\epsilon)
\left(1-\epsilon^2R(u)+O(\epsilon^3)\right)
\bigg] \frac{\delta(p_2^2-p_1^2)}{2\pi p}
\Pf\frac{1}{u} \\
    &= -2\gamma^2\gamma_{\rm E}
\frac{\delta(p_2^2-p_1^2)}{2\pi p}
\operatorname{Pf}\frac{1}{u}.
\label{eq:inverse-sine-finite}
\ee
Finally, the $\Pf \frac{\ln u}{u}$ part follows from the terms
proportional to $\epsilon$ in \cref{eq:pf-epsilon}. Their combined coefficient in \cref{eq:secondorder} is
\be
&\qquad \lim_{\epsilon \to 0^+} \gamma^2(2p\lambda)^{4\epsilon}
\bigg[
\Gamma(1-2\epsilon)\cos(\pi\epsilon)
\\
&\qquad\qquad -2\Gamma^2(1-2\epsilon)\cos^2(\pi\epsilon)
\left(1-\epsilon^2R(u)+O(\epsilon^3)\right)
\bigg] 
\frac{\delta(p_2^2-p_1^2)}{2\pi p}
\Pf\frac{\ln u}{u} \\
    &= -\gamma^2
\frac{\delta(p_2^2-p_1^2)}{2\pi p}
\Pf\frac{\ln u}{u}.
\label{eq:log-sine-finite}
\ee
Combining \cref{eq:diagonal-finite,eq:inverse-sine-finite,eq:log-sine-finite}, the second order amplitude is
\be
    \braket{ p_2|\op S^{(2)}| p_1 }
&=-\gamma^2\Bigg[
2\gamma_{\rm E}^2
\delta(\vec p_2-\vec p_1)+2\gamma_{\rm E}
\frac{\delta(p_2^2-p_1^2)}{2\pi p}
\Pf\frac{1}{\sin^2 \frac{\theta}{2}}
\\
&\qquad +\frac{\delta(p_2^2-p_1^2)}{2\pi p}
\Pf\frac{\ln \sin^2 \frac{\theta}{2}}{\sin^2\frac{\theta}{2}}
\Bigg].
\label{eq:secondorderamplitude}\ee
This is precisely the distributional second-order Coulomb amplitude, \cref{eq:S2-dist}.

%%---------------------------------------------------
%%=========================================================
\section{Heisenberg picture \& the LSZ formalism}
\label{sec:heisenberg-LSZ}

In the previous section, we have shown that the Coulomb scattering amplitude can be computed perturbatively in the Schr\"odinger picture, once the Dollard correction is appropriately taken into account. However, the Schr\"odinger picture evolution does not readily generalize to quantum field theory. With an eye towards applying these methods to infrared divergences in QFT, we reformulate our result in the Heisenberg picture.

In non-relativistic quantum mechanics, the available (time-dependent) operators are $(\vec{\op x}(t), \vec{\op p}(t))$ obtained from classical position and momenta $(\vec{x}(t), \vec{p}(t))$ along with their operator products. For evolution under suitably ``short range/short lived'' potentials, the classical variables $(\vec{x}(t), \vec{p}(t))$ and the quantum operators $(\vec{\op x}(t), \vec{\op p}(t))$ suitably asymptote to free evolution at large times. Then the Heisenberg S-matrix is a unitary operator relating the asymptotic operators at \(t \to -\infty\) to those at \(t \to +\infty\). Due to the long range nature of the static Coulomb potential, the large time classical trajectory of the particle does not approach a free solution but gets a logarithmic correction in time. This issue persists even in the quantum theory and obstructs the straightforward definition of the Heisenberg S-matrix. Below we show that Dollard's unitary correction essentially resolves this issue --- instead of correcting the evolution of states as in the Schr\"odinger picture, the Dollard correction gives us a canonical transformation to new classical variables which do asymptote to a free evolution. In the quantum picture, this is a time-dependent unitary transformation on the operators so that the new operators are asymptotically free and allow us to define the Heisenberg S-matrix. We show below that the Coulomb amplitudes can be computed directly in the Heisenberg picture, and also show how the LSZ formalism adapts to this case. We expect that this can be generalized to consider IR divergences in the QFT scattering amplitudes.\\

We first consider the classical theory with phase space variables \((\vec x(t), \vec p(t))\). The trajectory of a particle in a Coulomb potential at large times is \(t \to \infty\) is \cite{Reed-Simon3}
\be
    \vec x(t) \sim \frac{\vec p_\infty}{m} t - \frac{\alpha m \vec p_\infty}{p^3_\infty} \ln(4 H_0 t) + \vec x_\infty + O(\ln t/t) \eqsp \vec p \sim \vec p_\infty - \frac{\alpha m^2 \vec p_\infty}{p^3_\infty} \frac1t + O(\ln t/t^2) \,,
\label{eq:Coulomb-falloff}
\ee
where \((\vec x_\infty, \vec p_\infty)\) are constants of integration. It is apparent that at large time the Coulomb solution \(\vec x(t)\) does not asymptote to any free solution due to the logarithmic term in time.

We now define new variables \((\vec X(t), \vec P(t))\) which do asymptote to a free solution by a canonical transformation as follows. Let \(F_2\) be a generating function of type-II \cite{Goldstein:1950,Arnold:1989who,Landau:1975pou} given by 
\be
    F_2(\vec x,\vec P,t) \defn \vec x \cdot \vec P - \Phi(P,t) \,,
\ee
where \(\Phi(P,t)\) is as in \cref{eq:Phi-Dollard}. Then, the canonical change of variables is
\be
    \vec p &= \pd[F_2]{\vec x} = \vec P \,, \\
    \vec X &= \pd[F_2]{\vec P} = \vec x - \vec\nabla_P \Phi = \vec x - \alpha m \sgn(t) \frac{\vec P}{P^3} \lb( 2 - \ln\abs{4 H_0 t } \rb) \theta(\abs{4 H_0t} - 1 ) \,, \\
    K &= H + \pd[F_2]{t} = H_0 + V(x) - V_{\rm D}(p,t) \,,
\label{eq:XP-transform}
\ee
where \(K\) is the Hamiltonian for the time evolution of the new variables. Note that \(K\) should be expressed in terms of \((\vec X, \vec P)\) which makes the Coulomb potential \(V\) significantly more complicated than the usual Coulomb form; however we will not need this explicitly.

Using \cref{eq:Coulomb-falloff,eq:XP-transform} we find the asymptotic behaviour as
\be
    \vec X(t) \sim \frac{\vec P_\out}{m} t + \vec X_\out + O(\ln t/t) \eqsp \vec P(t) \sim \vec P_\out + O(1/t) \,,
\label{eq:XP-asymp}
\ee
where we have taken \(\vec X_\out = \vec x_\infty - 2 \alpha m \vec p_\infty/p_\infty^3\) and \(\vec P_\out = \vec p_\infty\). Thus, we can conclude that, as \(t \to \infty\), the evolution of the new variables \((\vec X(t), \vec P(t))\) asymptotes to the free solution \(\vec X_\out(t)\) with initial data \((\vec X_\out, \vec P_\out)\).

Now we implement this canonical transformation in the quantum theory as a unitary transformation. Consider the operator
\be
    \op{\mc W}(t) \defn e^{-i \Phi(\op p(t),t)} = \op U(t)^* \op W(t) \op U(t) \,.
\label{eq:scW-defn}
\ee
It follows from the analysis in \cite{Dollard:1964cok} that \(\op{\mc W}(t)\) is a unitary operator. The unitary transformation to the new operators \(\lb( \vec{\op X}(t), \vec{\op P}(t) \rb)\) is given by
\be
    \lb( \vec{\op X}(t), \vec{\op P}(t) \rb) &= \op{\mc W}(t) \lb( \vec{\op x}(t), \vec{\op p}(t) \rb) \op{\mc W}(t)^* \,.
\label{eq:xp-XP-unitary}
\ee
Noting that at \(t=0\), \(\lb(\vec{\op X}, \vec{\op P} \rb) = \lb(\vec{\op x}, \vec{\op p} \rb)\) and using \cref{eq:scW-defn} we get
\be
    \lb( \vec{\op X}(t), \vec{\op P}(t) \rb) = \op U(t)^* \op W(t) \lb( \vec{\op X}, \vec{\op P} \rb) \op{W}(t)^* \op U(t) \,.
\ee
Thus the unitary evolution operator \(\op{\mc U}(t)\) for the new variables is
\be
    \op{\mc U}(t) \defn \op W(t)^* \op U(t) = \op U(t) \op{\mc W}(t)^* \,,
\label{eq:scU-defn}
\ee
where the last equality follows from \cref{eq:scW-defn}. Here we note that \(\vec{\op P}(t) = \vec{\op p}(t)\), however the operators in \cref{eq:scU-defn} --- in particular \(\op U(t)\) --- are still written as functions of \(\vec{\op x}(t)\). For our analysis below we will not need to change all the operator variables to \((\vec{\op X}(t), \vec{\op P}(t))\).

In the Heisenberg picture, Dollard's definition of the M\o ller operators can be given as
\be
    \slim_{t \to \pm \infty} \op{\mc U}(t)^* \op U_0(t) = \slim_{t \to \pm \infty} \op U(t)^* \op W(t) \op U_0(t) = \slim_{t \to \pm \infty} \op U(t)^* \op U_{\rm D}(t) = \op\Omega_\pm \,.
\label{eq:mcU-Moller}
\ee
The time evolution of the new operators \(\lb( \op X(t), \op P(t) \rb)\) generated by \(\op{\mc U}(t)\) \emph{does} (strongly) asymptote to the free evolution; Dollard's modification to the free evolution is instead absorbed into the definition of these new operators.

Using the M\o ller operators we can define the asymptotic in/out operators by (see also \cite{Morchio:2014zga})
\be
    \lb( \vec{\op X}_{\out/\inn}, \vec{\op P}_{\out/\inn} \rb) &\defn \op\Omega_\pm \lb( \vec{\op X}, \vec{\op P} \rb) \op\Omega_\pm^*  = \op\Omega_\pm \lb( \vec{\op x}, \vec{\op p} \rb) \op\Omega_\pm^* \,
\label{eq:asymp-ops-defn}
\ee
whose time evolution is given by the free evolution with Hamiltonian
\be
    \op H_{\out/\inn} &\defn \op\Omega_\pm \op H_0(p) \op\Omega_\pm^* = \op\Omega_\pm \frac{\op p^2}{2m} \op\Omega_\pm^* = \frac{\op P_{\out/\inn}^2}{2m} \,.
\ee
Using \cref{eq:Moller-improper,eq:asymp-ops-defn} we have (in the sense of improper states)
\be
    \vec{\op P}_{\out/\inn} \ket{\varphi_k^\pm} &=  \op\Omega_\pm \vec{\op P} \op\Omega_\pm^* \ket{\varphi_k^\pm} = \op\Omega_\pm \vec{\op p} \ket{k} = \vec k~ \op\Omega_\pm \ket{k} = \vec k \ket{\varphi_k^\pm}
\label{eq:P-in-basis}
\ee
Thus, the Coulomb wavefunctions are exactly the improper eigenstates of the asymptotic momenta \(\vec{\op P}_{\out/\inn}\).

With this definition of the asymptotic free operators, we can define the Heisenberg S-matrix \(\op S_{\rm H}\) as a map from the algebra of in-operators to the algebra of out-operators. In terms of the asymptotic position and momentum operators, we have
\be
    \lb( \vec{\op X}_\out, \vec{\op P}_\out \rb) = \op S_{\rm H}^* \lb(\vec{\op X}_\inn, \vec{\op P}_\inn \rb) \op S_{\rm H}
\label{eq:SH-defn}
\ee
Using \cref{eq:asymp-ops-defn} we can express \(\op S_{\rm H}\) in terms of the M\o ller operators as follows:
\be
    \vec{\op X}_\out &= \op\Omega_+ \vec{\op X} \op\Omega_+^* = \op\Omega_+ \op\Omega_-^* \vec{\op X}_\inn \op\Omega_- \op\Omega_+^* \\
    \implies \op S_{\rm H} &=  \op\Omega_- \op\Omega_+^* \,.
\label{eq:SH-Moller}
\ee
From \cref{eq:S-defn}, the Schr\"odinger and Heisenberg S-matrices are related through the M\o ller operators by \cite{Morchio:2014zga}
\be
    \op S_{\rm H} = \op\Omega_- \op S \op\Omega_-^* = \op\Omega_+ \op S \op\Omega_+^* \,.
\label{eq:S-SH}
\ee
Further, using \cref{eq:Moller-basis} we get the basis expression
\be
    \op S_{\rm H} &= \int d^3k d^3k'~ \ket{\varphi_k^-}\braket{k | k'} \bra{\varphi_{k'}^+} = \int d^3k~ \Dop{\varphi_k^-}{\varphi_k^+} \,.
\label{eq:SH-basis}
\ee
Using this expression it follows that the matrix elements of \(\op S_{\rm H}\) in the in (or out) Coulomb basis gives the scattering amplitudes
\be
    \braket{\varphi_k^- | \op S_{\rm H} | \varphi_{k'}^-} &=  \braket{\varphi_k^+ | \op S_{\rm H} | \varphi_{k'}^+} = \braket{k | \op S | k'} = \braket{\varphi_k^+ | \varphi_{k'}^-}
\ee
So the Coulomb scattering amplitudes can be computed either
\begin{enumerate*}
    \item as the matrix elements of the Schr\"odinger S-matrix \(\op S\) in the momentum eigenbasis \(\ket{k}\), or
    \item as the matrix elements of the Heisenberg S-matrix \(\op S_{\rm H}\) in the incoming (or outgoing) Coulomb wavefunction basis \(\ket{\varphi^-_k}\), which are also the eigenbasis of the asymptotic momentum \(\vec{\op P}_\inn\).
\end{enumerate*} Now we obtain a formula for the scattering amplitudes directly using the Heisenberg S-matrix and the asymptotic in-operators in a few different ways.

From \cref{eq:asymp-ops-defn}, it follows that the M\o ller operator \(\op\Omega_-\) implements a unitary transformation which replaces the arguments of each operator from \((\op x, \op p)\) to \((\op X_\inn, \op P_\inn)\). Thus, we have
\be
    \op W_\inn(t) &\defn \op\Omega_- \op W(t) \op\Omega_-^* = \op\Omega_- e^{- i \Phi (\op p, t)} \op\Omega_-^* = e^{- i \Phi (\op P_\inn, t)} \\
    \op V_{\rm I, \inn}(t) &\defn \op\Omega_- \op V_{\rm I}(t) \op\Omega_-^* = \op U_\inn(t)^* \op V(X_\inn) \op U_\inn(t)
\label{eq:in-conversion}
\ee
where we have used the subscript ``\(\inn\)'' to denote that the operators are computed using the same expressions but with the in-operators. Then, using \cref{eq:perturbativeSmatrix} for the Schr\"odinger S-matrix in \cref{eq:S-SH} and inserting \(\1 = \op\Omega_- \op\Omega_-^*\) between the operators, we get
\be
    \op S_{\rm H} = \op W_{\inn,\epsilon}(+\infty)^*  \Tord{ \exp\lb( - i \int_{-\infty}^{+\infty} dt'~ \op V_{\rm I, \inn, \epsilon}(t') \rb) } \op W_{\inn, \epsilon}(-\infty) \,,
\label{eq:SH-formula}
\ee
where on the right-hand side we have reintroduced our analytic regulator \(\epsilon\) (see \cref{sec:Born-series}) since the individual operators do not have a large time limit by themselves, even though the full right-hand side does.

To explicitly compute the scattering amplitudes in the Heisenberg picture, we only have to note that the Heisenberg S-matrix (\cref{eq:SH-formula}) is identical to the Schr\"odinger one (\cref{eq:perturbativeSmatrix}) except written in terms of the in-operators and that the Coulomb in-wavefunctions are eigenstates of \(\vec{\op P}_\inn\) (\cref{eq:P-in-basis}). Thus, the computation of the amplitude \(\braket{\varphi_k^- | \op S_{\rm H} | \varphi_{k'}^-}\) proceeds exactly as in \cref{sec:Born-series}.

We can also compute \(\op S_{\rm H}\) by directly relating the $\lb( \op X(t), \op P(t) \rb)$ to the free solutions $\lb( \op X_\inn(t), \op P_\inn(t) \rb)$ by an ``in-interaction picture'' as follows. Using \cref{eq:asymp-ops-defn} in the evolution of \(\op X(t)\) we get
\be
    \op X(t) &= \op{ \mc U}(t)^* \op X \op{\mc U}(t) 
    = \op{ \mc U}(t)^* \op\Omega_-^* \op X_\inn \op\Omega_- \op{\mc U}(t) \\
    &= \op{ \mc U}(t)^* \op\Omega_-^* \op U_\inn(t) \op X_\inn(t) \op U_\inn(t)^* \op\Omega_- \op{\mc U}(t) \\
    &= \op{ \mc U}(t)^* \op U_0(t) \op\Omega_-^* \op X_\inn(t) \op\Omega_- \op U_0(t)^* \op{\mc U}(t) \\
    &= \op{ \mc U}_{\rm I, \inn}(t)^* \op X_\inn(t) \op{\mc U}_{\rm I, \inn}(t)
\label{eq:X-Xin-int-pic}
\ee
where we have defined
\be
    \op{\mc U}_{\rm I, \inn}(t) \defn \op \Omega _- \op U_0(t)^* \op{\mc U}(t) = \op U_\inn(t)^* \op \Omega _- \op{\mc U}(t) \eqsp \op{\mc U}_{\rm I, \inn}(0) = \op \Omega_- \,.
\label{eq:scU-int-defn}
\ee
Further from \cref{eq:mcU-Moller,eq:SH-Moller} we also see that
\be
    \slim_{t \to -\infty} \op{\mc U}_{\rm I, \inn}(t) = \op\Omega_- \op\Omega_-^* = \1 \eqsp \slim_{t \to +\infty} \op{\mc U}_{\rm I, \inn}(t) = \op\Omega_- \op\Omega_+^* = \op S_{\rm H}
\ee
Thus, as \(t \to -\infty\) the \(\op{\mc U}_{\rm I, \inn}(t)\) matches the operator \(\op X(t)\) to \(\op X_\inn(t)\) while as \(t \to +\infty\) it matches the out-operator through the Heisenberg S-matrix. To obtain an explicit formula for \(\op{\mc U}_{\rm I, \inn}(t)\) and hence \(\op S_{\rm H}\), we rewrite \(\op{\mc U}_{\rm I, \inn}(t)\) in a convenient form as follows:
\be
    \op{\mc U}_{\rm I, \inn}(t) &= \op \Omega _- \op U_0(t)^* \op{\mc U}(t) = \op \Omega _- \op U_0(t)^* \op W(t)^* \op U(t) \\
    &= \op \Omega _- \op W(t)^* \op U_0(t)^* \op U(t) = \op W_\inn(t)^* \op \Omega _- \op U_0(t)^* \op U(t) \\
    &= \op W_\inn(t)^* \op \Omega _- \op U_{\rm I}(t)
\ee
where in the first line we have used \cref{eq:scU-defn,eq:scU-int-defn}, in the second line the fact that \(\op W(t)\) commutes with \(\op U_0(t)\) and \cref{eq:in-conversion}, and finally in the last line the ``interaction picture unitary'' \(\op U_{\rm I} \defn \op U_0(t)^* \op U(t)\). From this it is straightforward to obtain the differential equation
\be
    i \td{t} \big[ \op W_\inn(t) \op{\mc U}_{\rm I, \inn}(t) \big] = \op\Omega_- \op V_{\rm I}(t) \op U_{\rm I}(t) 
    = \op V_{\rm I,\inn}(t) \big[ \op W_\inn(t) \op{\mc U}_{\rm I,\inn}(t) \big] \,,
\ee
which can be solved with the asymptotic boundary condition \(\op{\mc U}_{\rm I, \inn}(-\infty) = \1\) to get
\be
    \op{\mc U}_{\rm I, \inn}(t) &=  \op W_{\inn, \epsilon}(t)^* \Tord \exp\lb(-i\int_{-\infty}^{t}\,dt'~ \op V_{\rm I, \inn,\epsilon}(t') \rb) \op W_{\inn, \epsilon}(-\infty) \,,
\label{eq:mcU-in-soln}
\ee
where on the right-hand side we have reintroduced our analytic regulator \(\epsilon\) (see \cref{sec:Born-series}) since the individual operators do not have a limit to large times. Taking the limit \(t \to +\infty\), we get \(\op S_{\rm H}\) which reproduces \cref{eq:SH-formula}.\\

%%---------------------------------------------------------
\subsection{LSZ formalism}
\label{sec:LSZ}

Next, we show that the Heisenberg S-matrix can also be computed directly by closely following the LSZ formalism \cite{Lehmann:1954rq} used in QFT. We first recall the classical constructions required and then formulate the LSZ asymptotic conditions relevant for quantum mechanics. We argue, but do not prove, that the new operators defined above satisfy these conditions and then derive the scattering amplitudes via the analogue of the LSZ reduction formula.\\

We consider the phase space of solutions \(\vec x(t)\) in classical mechanics with the \emph{symplectic form}
\be
    \Omega\lb(\delta_1 \vec x(t), \delta_2 \vec x(t)\rb) &= m \lb[ \delta_1 \vec x(t) \cdot \td{t} \delta_2 \vec x(t) - \td{t} \delta_1 \vec x(t) \cdot \delta_2 \vec x(t) \rb] \\
    &= \delta_1 \vec x(t) \cdot \delta_2 \vec p(t) - \delta_1 \vec p(t) \cdot \delta_2 \vec x(t)
\label{eq:symp-form}
\ee
where \(\delta \vec x(t)\) denotes the variation of the solution; for the free particle the solution space is linear, we can drop the \(\delta\). Taking the time derivative
\be
    \td{t} \Omega\lb(\delta_1 \vec x(t), \delta_2 \vec x(t)\rb) &= m \lb[ \delta_1 \vec x(t) \cdot \frac{d^2}{dt^2} \delta_2 \vec x(t) - \frac{d^2}{dt^2} \delta_1 \vec x(t) \cdot \delta_2 \vec x(t) \rb]
\label{eq:symp-form-dt}
\ee
Thus, the symplectic form is conserved if
\begin{enumerate*}
    \item both \(\delta_1 \vec x(t)\) and \(\delta_2 \vec x(t)\) are linearizations of the free solutions (or both full free solutions), or
    \item both \(\delta_1 \vec x(t)\) and \(\delta_2 \vec x(t)\) are linearizations of the Coulomb solutions; or any other fixed potential.
\end{enumerate*}

Using \(\vec x \equiv x^i\) and the canonical momentum \(\vec p \equiv p_i\) as the canonical coordinates on phase space, the symplectic form is usually written as a \(2\)-form on phase space $d x^i \wedge d p_i$ (see \cite{Arnold:1989who}). Then, \cref{eq:symp-form} is this \(2\)-form contracted into two tangent vector fields on phase space. From the transformation \cref{eq:XP-transform} it can be verified that \(d x^i \wedge dp_i = d X^i \wedge dP_i\), that is, the transformation preserves the symplectic form and hence is canonical.\\

In the LSZ formalism \cite{Lehmann:1954rq}, we need to consider the symplectic product \(\Omega\lb(\vec x(t), \vec x_\free(t)\rb)\) where \(\vec x_\free(t) = \vec p_\free t/m + \vec x_\free\) is a free solution with initial data \(\lb(\vec x_\free, \vec p_\free\rb)\) while \(\vec x(t)\) is an interacting solution. We emphasize that the full nonlinear solution \(\vec x(t)\) is used here instead of the linearized solution \(\delta \vec x(t)\). The solution \(\vec x(t)\) asymptotes to a free solution \(x_\inn(t)\) as \(t \to -\infty\) in the sense of LSZ if
\be
    \lim_{t \to -\infty} \Omega\lb(\vec x(t), \vec x_\free(t)\rb) = \Omega\lb(\vec x_\inn(t), \vec x_\free(t)\rb)
\label{eq:LSZ-limit-classical}
\ee
for all free solutions \(\vec x_\free(t)\). Note that since both \(\vec x_\inn(t)\) and \(\vec x_\free(t)\) are free solutions the right-hand side is time-independent. For ``short range/short lived'' potentials, the interacting solutions \(\vec x(t)\) decay at large times and \cref{eq:LSZ-limit-classical} holds and defines the asymptotic in-solution \(x_\inn(t)\) in the LSZ sense. Thus, the asymptotic phase space is the one for a free theory and can be quantized.

But, for the Coulomb potential \cref{eq:LSZ-limit-classical} fails since the left-hand side does not have a finite limit due to the slower falloff in time (\cref{eq:Coulomb-falloff}) of the solutions \(\vec x(t)\). Instead we must consider \cref{eq:LSZ-limit-classical} with the Coulomb solutions \(\vec X(t)\) in the transformed variables\footnote{As has been emphasized in \cite{Schwartz:2014sze}, the LSZ formalism does not require us to use the ``elementary'' fields which directly appear in the Lagrangian.} \cref{eq:XP-transform}. Using \cref{eq:XP-asymp}, as \(t \to -\infty\) we get
\be
    \Omega\lb(\vec X(t), \vec x_\free(t)\rb) &= m \lb[ \vec X(t) \cdot \td{t} \vec x_\free - \td{t} \vec X(t) \cdot \vec x_\free(t) \rb] \\
    &= \lb( \frac{\vec P_\inn}{m} t + \vec X_\inn \rb) \cdot \vec p_\free - \vec P_\inn \cdot \lb( \frac{\vec p_\free}{m} t + \vec x_\free \rb) + O(\ln t/t) \\
    &= \vec X_\inn \cdot \vec p_\free - \vec P_\inn \cdot \vec x_\free + O(\ln t/t) \,.
\ee
This implies that
\be
    \lim_{t \to -\infty} \Omega\lb(\vec X(t), \vec x_\free(t)\rb) &= \Omega\lb(\vec X_\inn(t), \vec x_\free(t)\rb) = \vec X_\inn \cdot \vec p_\free - \vec P_\inn \cdot \vec x_\free \,,
\label{eq:LSZ-asymp-X}
\ee
where, as before, \(\vec X_\inn(t)\) is the free solution with initial data \((\vec X_\inn, \vec P_\inn)\). Thus, the LSZ asymptotic condition (\cref{eq:LSZ-limit-classical}) holds for the classical Coulomb solutions in terms of the new modified variables \(\vec X(t)\). An exactly analogous computation also shows that the future asymptotic limit (as \(t \to + \infty\)) holds with the asymptotic field \(\vec X_\out(t)\). \\

In the LSZ approach to quantum scattering we require that the analogue of the asymptotic limits in \cref{eq:LSZ-asymp-X} also hold at the operator level in the following sense (see \cite{Lehmann:1954rq}). For the Heisenberg-picture operator \(\vec{\op X}(t)\) there exists a free evolving operator \(\vec{\op X}_\inn(t)\) such that
\be
    \lim_{t \to -\infty} \Omega\big( \vec{\op X}(t) - \vec{\op X}_\inn(t), \vec x_\free(t) \big) = 0 \,.
\label{eq:LSZ-limit}
\ee
Using \cref{eq:symp-form} and considering the two cases \(\vec p_\free = 0\) with \(\vec x_\free(t) = \vec x_\free\), and \(\vec x_\free = 0\) with \(\vec x_\free(t) = \vec p_\free t/m\), this is equivalent to requiring
\be
    \lim_{t \to -\infty} \lb[ m \td{t} \vec{\op X}(t) - \vec{\op P}_\inn \rb] = 0 \eqsp \lim_{t \to -\infty} \lb[ \vec{\op X}(t) - t \td{t} \vec{\op X}(t) - \vec{\op X}_\inn \rb] = 0 \,.
\label{eq:LSZ-limit-2}
\ee
The proof of these LSZ asymptotic conditions does not follow directly from the existence of Dollard's modified M\o ller operator \(\op\Omega_-\), since these are statements about the limits of unbounded operators including the time-derivatives of \(\vec{\op X}(t)\). We do not attempt a proof here however, we expect that a proof can be given along the following general lines: As in \cite{Dollard:1964cok}, consider the action of these operators on a state \(\ket{\psi} \in \Hilb\) such that the momentum-space wavefunction \(\psi(\vec p)\) is compactly supported away from \(\vec p = 0\); such states form a dense subspace in \(\Hilb\). Then, using Cook's method and decay estimate techniques in \cite{Dollard:1964cok, Reed-Simon3}, we expect that the suitable limits in \cref{eq:LSZ-limit,eq:LSZ-limit-2} can be shown on this dense subspace.

In the following we assume the LSZ asymptotic limits hold, in both limits \(t \to \pm \infty\), for \(\vec{\op X}(t)\) and obtain \(\op S_{\rm H}\) by directly comparing the asymptotic in- and out-operators through the free equation of motion, similar to the LSZ reduction formula.

From \cref{eq:symp-form-dt}, since \(\vec x_\free(t)\) is a free solution we get
\be
    \td{t} \Omega\lb( \vec{\op X}(t), \vec x_\free(t) \rb) = - \frac{d^2}{dt^2} \vec{\op X}(t) \cdot \vec x_\free(t) \,.
\ee
Choosing the free solution to be \(p_\free = 0\) and $\vec x_\free(t) = \vec x_\free$ in \cref{eq:LSZ-limit} and integrating from $t \to -\infty$ to $t \to +\infty$, we get
\be
    \vec{\op P}_\out - \vec{\op P}_\inn = m \int_{-\infty}^{+\infty}dt \frac{d^2}{dt^2} \vec{\op X}(t) \,,
\label{eq:LSZ-red}
\ee
where we have used the LSZ asymptotic conditions \cref{eq:LSZ-limit-2}. Thus, the difference of the in- and out- operators can be expressed in terms of the free equation of motion acting on the full interacting operator \(\vec{\op X}(t)\). This is the direct analogue in quantum mechanics of the \emph{LSZ reduction formula}.

In QFT the LSZ reduction formula expresses the scattering amplitudes in terms of the vacuum expected values of the interacting field. In our quantum mechanics case, we can evaluate this more directly. Using \cref{eq:X-Xin-int-pic,eq:mcU-in-soln}, the LSZ asymptotic limit \(m~ d\vec{\op X}/dt \to \vec{\op P}_\inn \) (\cref{eq:LSZ-limit-2}) and \(\op{\mc U}_{\rm I, \inn}(-\infty) = \1\) we get
\be
    \vec{\op P}_\out - \vec{\op P}_\inn 
    &= m \td{t} \vec{\op X}(t) \bigg\vert_{t = -\infty}^{t = + \infty} 
    = \op{\mc U}_{\rm I, \inn}(t)^* \vec{\op P}_\inn \op{\mc U}_{\rm I, \inn}(t) \bigg\vert_{t = -\infty}^{t = + \infty} \\[2ex]
    \implies \vec{\op P}_\out &= \op{\mc U}_{\rm I, \inn}(+\infty)^* \vec{\op P}_\inn \op{\mc U}_{\rm I, \inn}(+\infty)
\ee
Comparing with the definition of \(\op S_{\rm H}\) (\cref{eq:SH-defn}) and using \cref{eq:mcU-in-soln} we again reproduce \cref{eq:SH-formula}. We remark here that since the LSZ asymptotic limits are assumed to hold only on a dense subspace of the Hilbert space, the \(\op S_{\rm H}\) defined by the above procedure is also defined only on this dense subspace. However, it was shown by Herbst \cite{Herbst:1974hi} that this S-matrix can be uniquely extended to the full Hilbert space as a unitary operator.

%%+++++++++++++++++++++++++++++++++++++++++++++++++++++++++++++++
\appendix

\section{Conventions \& useful identities}
\label{sec:conventions}

Throughout this paper, we use units with \(\hbar = 1\). We also denote quantum observables/operators by the boldfaced version of the symbol for the corresponding classical observable; e.g., the quantum operator corresponding to the classical position \(\vec{x}\) is denoted by \(\vec{\op x}\).

We use the symmetric Fourier transform conventions:
\be
    f(\vec x) &= \frac{1}{(2\pi)^{\nfrac32}} \int d^3q~ f(\vec q) e^{-i \vec q \cdot \vec x} \eqsp
    f(\vec q) &= \frac{1}{(2\pi)^{\nfrac32}} \int d^3x~ f(\vec x) e^{i \vec q \cdot \vec x}
\ee
and, for the time coordinate,
\be
    f(t) &= \frac{1}{(2\pi)^\half} \int d\omega~ f(\omega) e^{-i \omega t} \eqsp
    f(\omega) &= \frac{1}{(2\pi)^\half} \int dt~ f(t) e^{i \omega t}.
\ee
With these conventions, the Parseval-Plancherel theorem takes the form
\be
    \int d^3x~ \bar f_1(\vec x) f_2(\vec x)
    = \int d^3q~ \bar f_1(\vec q) f_2(\vec q).
\label{eq:PP}
\ee
We will also make use of the following distributional Fourier transforms:
\be
    \frac{1}{(2\pi)^\half} \int_0^{+\infty} dt~ e^{i \omega t}
    = \frac{1}{(2\pi)^\half} \frac{i}{\omega + i0^+} \eqsp
    \frac{1}{(2\pi)^\half} \int_{-\infty}^0 dt~ e^{i \omega t}
    = \frac{1}{(2\pi)^\half} \frac{-i}{\omega - i0^+}.
\label{eq:FT-step}
\ee
Consequently, using \cref{eq:PP}, we obtain the half-line integral identities
\be
    \int_0^{+\infty} dt~ f(t)
    = \frac{1}{(2\pi)^\half} \int d\omega~ \frac{-i}{\omega - i0^+} f(\omega) \eqsp
    \int_{-\infty}^0 dt~ f(t)
    = \frac{1}{(2\pi)^\half} \int d\omega~ \frac{+i}{\omega + i0^+} f(\omega).
\label{eq:FT-half}
\ee
The above conventions also determine the normalization of the momentum eigenstates used throughout the paper. To see this, consider a state \(\ket{\psi}\) in \(L^2(\bb R^3)\), with position-space wavefunction \(\psi(\vec x)=\braket{\vec x|\psi}\). We define its momentum-space wavefunction by
\be
    \braket{\vec q|\psi} \defn \psi(\vec q)
    = \frac{1}{(2\pi)^{\nfrac32}} \int d^3x~ \psi(\vec x) e^{i\vec q\cdot\vec x}.
\ee
It follows that the momentum eigenstates have position-space wavefunctions
\be
    \braket{\vec x|\vec q}
    = \frac{1}{(2\pi)^{\nfrac32}}e^{-i\vec q\cdot\vec x}.
\ee
The inner product can therefore be expressed in either representation as
\be
    \braket{\psi_1|\psi_2}
    = \int d^3q~ \bar\psi_1(\vec q)\psi_2(\vec q)
    = \int d^3q~ \braket{\psi_1|\vec q}\braket{\vec q|\psi_2}.
\ee
In particular, the momentum eigenstates satisfy the resolution of the identity
\be
    \1 = \int d^3q~ \Dop{\vec q}{\vec q} \eqsp
    \braket{\vec q_1|\vec q_2} = \delta(\vec q_1-\vec q_2).
\label{eq:1-resolve}
\ee

Below we also collect some distributional identities used in the main paper. For \(x \in \bb R\), the \emph{Sokhotski-Plemelj} formula gives
\be
    \frac{1}{x \pm i0^+} = \pv \frac1x \mp i\pi \delta(x).
\label{eq:SP}
\ee
We also use analytic regularization to define finite-part distributions following \cite{Gelfand1}. For \(x \in [0,1]\), the \emph{Hadamard finite part} (or \emph{partie finite}) distribution is defined by
\be
    \int_0^1 dx~ \left(\Pf \frac{1}{x}\right) f(x)
    &\defn \lim_{\epsilon \to 0^+}
    \left[
    \int_0^1 dx~ \frac{f(x)}{x^{1-\epsilon}}
    - \frac{f(0)}{\epsilon}
    \right] = \lim_{\epsilon \to 0^+}
    \int_0^1 dx~
    \left(\frac{1}{x^{1-\epsilon}}-\frac{\delta(x)}{\epsilon}\right)f(x).
\ee
where \(f(x)\) is a smooth test function. Thus, in the sense of distributions we have
\be
    \Pf \frac{1}{x} = \lim_{\epsilon \to 0^+}     \left(\frac{1}{x^{1-\epsilon}} - \frac{\delta(x)}{\epsilon}\right) 
\label{eq:Pf-x}
\ee
More generally,
\be
    \Pf \frac{(\ln x)^n}{x}
    &\defn \lim_{\epsilon \to 0^+}
    \frac{1}{n!}\frac{d^n}{d\epsilon^n}
    \left[\frac{1}{x^{1-\epsilon}} - \frac{\delta(x)}{\epsilon}\right] = \lim_{\epsilon \to 0^+}
    \left[
    \frac{1}{n!} \frac{(\ln x)^n}{x^{1-\epsilon}}
    -\frac{(-1)^n\delta(x)}{\epsilon^{n+1}}
    \right] \,,
\label{eq:Pf-ln}
\ee
which also gives the distributional Laurent series
\be
    \frac{1}{x^{1-\epsilon}} = \frac{\delta(x)}{\epsilon} + \sum_{n=0}^\infty \epsilon^n \Pf \frac{(\ln x)^n}{x} \,.
\label{eq:series}
\ee

%%====================================================
\section{Distributional scattering amplitude \& perturbative expansion}
\label{sec:Herbst}

It was shown by Herbst \cite{Herbst:1974hi} that the na\"ive expression for the amplitude \cref{eq:amplitude-naive} is more correctly understood as a distribution acting on a dense subspace of smooth, and square-integrable wavefunctions \(\psi(\vec k)\) given by
\be
    (\op S \psi)(\vec k) &= \int d^3k'~ S(\vec k, \vec k') \psi(\vec k') \\
    &= \lim_{\epsilon \to 0^+} \frac{\Gamma(1+i\gamma)}{\Gamma(1-i\gamma)} \frac{\gamma}{2\pi i k} \int d^3 k'~ \delta(k^2 - {k'}^2) \lb( \sin^2 \tfrac{\theta}{2} \rb)^{-1-i\gamma + \epsilon} \psi(\vec k')
\label{eq:Herbst-dist-amplitude}
\ee
where \(\theta\) is the angle between \(\vec k\) and \(\vec k'\). Adding and subtracting \(\psi(\vec k)\) on the right-hand side we get
\be
    (\op S \psi)(\vec k) &= \frac{\Gamma(1+i\gamma)}{\Gamma(1-i\gamma)} \Bigg[ \psi(\vec k) + \frac{\gamma}{2\pi i k} \lim_{\epsilon \to 0^+} \int d^3 k'~ \delta(k^2 - {k'}^2) \lb( \sin^2 \tfrac{\theta}{2} \rb)^{-1-i\gamma + \epsilon} \lb( \psi(\vec k') - \psi(\vec k) \rb) \Bigg]
\label{eq:Herbst-dist}
\ee
where we have safely taken the \(\epsilon \to 0^+\) in the first term. Note that the integral in the second term is now finite in the forward direction, \(\theta = 0\). Thus, in this form we can safely take the \(\gamma \to 0\) limit and obtain the expected free scattering amplitude.\footnote{In \cref{eq:Herbst-dist-amplitude} the na\"ive \(\gamma \to 0\) limit gives zero. However, the limits in \(\gamma\) and \(\epsilon\) cannot be interchanged, see also \cite{Herbst:1974hi}.  }

With the integral regularized we get
\be
    (\op S \psi)(\vec k) &= \frac{\Gamma(1+i\gamma)}{\Gamma(1-i\gamma)} \Bigg[ \psi(\vec k) + \frac{\gamma}{2\pi i k} \int d^3 k'~ \delta(k^2 - {k'}^2) \lb( \sin^2 \tfrac{\theta}{2} \rb)^{-1-i\gamma} \bigg( \psi(\vec k') - \psi(\vec k) \bigg) \Bigg] \\
    &= \frac{\Gamma(1+i\gamma)}{\Gamma(1-i\gamma)} \Bigg[ \psi(\vec k) + \frac{\gamma}{2\pi i k} \int d^3 k'~ \delta(k^2 - {k'}^2) \Pf \lb( \sin^2 \tfrac{\theta}{2} \rb)^{-1-i\gamma} \psi(\vec k') \Bigg] \,,
\label{eq:Herbst-dist1}
\ee
where \(\Pf\) is the Hadamard finite part distribution. Thus, we get the distributional amplitudes
\be
    S(\vec k, \vec k') = \frac{\Gamma(1+i\gamma)}{\Gamma(1-i\gamma)} \Bigg[ \delta(\vec k - \vec k') + \frac{\gamma}{2\pi i k} \delta(k^2 - {k'}^2) \Pf \lb( \sin^2 \tfrac{\theta}{2} \rb)^{-1-i\gamma} \Bigg] \,.
\label{eq:amplitude-dist}
\ee\\

Now we obtain a perturbative expansion of this distributional amplitude in the coupling constant \(\alpha\), equivalently \(\gamma\). Even though the amplitude is not an ordinary function, it is apparent that the left-hand side of \cref{eq:Herbst-dist1} depends analytically on \(\gamma\), and can be expanded into a series where each term in the expansion is itself a well-defined distribution integrated with \(\psi(\vec k')\), see \cite{Gelfand1} for details.

To obtain this perturbative series we first note the series
\be
    \lb( \sin^2 \tfrac{\theta}{2} \rb)^{-1-i\gamma + \epsilon} &= \lb( \sin^2 \tfrac{\theta}{2} \rb)^{-1  + \epsilon} \sum_{n=0}^\infty \frac{(-i\gamma)^n}{n!} \lb( \ln \sin^2 \tfrac{\theta}{2} \rb)^n \,.
\ee
We can use this expansion, with \(\epsilon = 0\), directly in \cref{eq:amplitude-dist} to obtain the required series in terms of \(\Pf \lb[ \lb( \sin^2 \tfrac{\theta}{2} \rb)^{-1} (\ln \sin^2 \tfrac{\theta}{2})^n \rb]\). However, to compare with our analysis in \cref{sec:Born-series} it will be more useful to start from \cref{eq:Herbst-dist} with a non-zero \(\epsilon\). We start with the ``subtraction term'', taking \(u = \sin^2\tfrac{\theta}{2}\),
\be
    &\quad \int d^3 k'~ \delta(k^2 - {k'}^2) \lb( \sin^2 \tfrac{\theta}{2} \rb)^{-1-i\gamma + \epsilon} \psi(\vec k) \\[1.5ex]
    &= \psi(\vec k) \sum_{n=0}^\infty \frac{(-i\gamma)^n}{n!} \int d^3 k'~ \delta(k^2 - {k'}^2) \lb( \sin^2 \tfrac{\theta}{2} \rb)^{-1  + \epsilon}  \lb( \ln \sin^2 \tfrac{\theta}{2} \rb)^n \\
    &= \psi(\vec k)~ 2\pi k \sum_{n=0}^\infty \frac{(-i\gamma)^n}{n!} \int_0^1 du~ u^{-1  + \epsilon}  \lb( \ln u \rb)^n 
    = \psi(\vec k)~ 2\pi k \sum_{n=0}^\infty \frac{(-i\gamma)^n}{n!} \frac{\partial^n}{\partial \epsilon^n}\int_0^1 du~ u^{-1  + \epsilon} \\
    &= \psi(\vec k)~ 2\pi k \sum_{n=0}^\infty \frac{(i\gamma)^n}{\epsilon^{n+1}} 
    = 2\pi k \sum_{n=0}^\infty \frac{(i\gamma)^n}{\epsilon^{n+1}} \int d^3k'~ \delta(\vec k - \vec k') \psi(\vec k')
\ee
With this \cref{eq:Herbst-dist} becomes
\be
    (\op S \psi)(\vec k) &= \frac{\Gamma(1+i\gamma)}{\Gamma(1-i\gamma)} \int d^3 k'~  \psi(\vec k') \Bigg[ \delta(\vec k - \vec k') \\
    &\quad + \lim_{\epsilon \to 0^+} \sum_{n=1}^\infty (i\gamma)^n \lb( \frac{\delta(\vec k - \vec k')}{\epsilon^n} + \frac{(- 1)^n}{(n-1)!} \frac{\delta(k^2 - {k'}^2)}{2\pi k}  \frac{ \lb( \ln \sin^2 \tfrac{\theta}{2} \rb)^{n-1} }{ \lb( \sin^2 \tfrac{\theta}{2} \rb)^{1 - \epsilon} } \rb) \Bigg]
\label{eq:Herbst-dist2}
\ee
which gives the distributional expansion of the amplitude
\be
    S(\vec k, \vec k') &= \frac{\Gamma(1+i\gamma)}{\Gamma(1-i\gamma)} \Bigg[ \delta(\vec k - \vec k') \\
    &\quad + \lim_{\epsilon \to 0^+} \sum_{n=1}^\infty (i\gamma)^n \lb( \frac{\delta(\vec k - \vec k')}{\epsilon^n} + \frac{(- 1)^n}{(n-1)!} \frac{\delta(k^2 - {k'}^2)}{2\pi k}  \frac{ \lb( \ln \sin^2 \tfrac{\theta}{2} \rb)^{n-1} }{ \lb( \sin^2 \tfrac{\theta}{2} \rb)^{1 - \epsilon} } \rb) \Bigg] \\[2ex]
    &= \frac{\Gamma(1+i\gamma)}{\Gamma(1-i\gamma)} \Bigg[ \delta(\vec k - \vec k') 
    + \sum_{n=1}^\infty \frac{(- i\gamma)^n}{(n-1)!} \frac{\delta(k^2 - {k'}^2)}{2\pi k}  \Pf \frac{ \lb( \ln \sin^2 \tfrac{\theta}{2} \rb)^{n-1} }{ \sin^2 \tfrac{\theta}{2} } \Bigg]
\ee
where the limit \(\epsilon \to 0^+\) is taken in the sense of distributions, that is, \emph{after} integrating with an appropriate test function. In the last step we have used \(\delta(\vec k - \vec k') = \frac{\delta(k^2 - {k'}^2)}{2 \pi k} \delta (\sin^2\theta/2)\) and \cref{eq:Pf-ln}.

To get the full expansion in powers of the coupling constant \(\alpha\) we also use
\be
    \frac{\Gamma(1+i\gamma)}{\Gamma(1-i\gamma)} &= 1 - 2i \gamma_{\rm E} \gamma - 2 \gamma_{\rm E}^2 \gamma^2 + O(\gamma^3)
\ee
where \(\gamma_{\rm E}\) is the Euler-Mascheroni constant. Then, the first few terms are given by
\be
    S^{(0)}(\vec k, \vec k') = \delta(\vec k - \vec k') \,,
\label{eq:S0-dist}
\ee
which is the free scattering amplitude as expected. We also get
\be
    S^{(1)}(\vec k, \vec k') &= - i \gamma \lb[ 2 \gamma_{\rm E} \delta(\vec k-\vec k') - \lim_{\epsilon \to 0^+} \lb(\frac{\delta(\vec k-\vec k')}{\epsilon} - \frac{ \delta(k^2-{k'}^2) }{2\pi k} \lb( \sin^2\tfrac{\theta}{2} \rb)^{-1+\epsilon} \rb) \rb] \\
    &= - i \gamma \lb[ 2 \gamma_{\rm E} \delta(\vec k-\vec k') + \frac{ \delta(k^2-{k'}^2) }{2\pi k} \Pf \lb( \sin^2\tfrac{\theta}{2} \rb)^{-1} \rb] \,,
\label{eq:S1-dist}
\ee
and
\be
    S^{(2)}(\vec k, \vec k') &= \gamma^2 \Bigg[ - 2\gamma_{\rm E}^2 \delta(\vec k-\vec k') + 2\gamma_{\rm E} \lim_{\epsilon \to 0^+} \lb( \frac{\delta(\vec k-\vec k')}{\epsilon} - \frac{\delta(k^2-{k'}^2)}{2\pi k} \lb( \sin^2\tfrac{\theta}{2} \rb)^{-1+\epsilon} \rb) \\
    &\qquad\qquad - \lim_{\epsilon \to 0^+} \lb( \frac{\delta(\vec k-\vec k')}{\epsilon^2} + \frac{\delta(k^2-{k'}^2)}{2\pi k} \frac{ \ln\sin^2\tfrac{\theta}{2} }{ \lb( \sin^2\tfrac{\theta}{2} \rb)^{1-\epsilon} } \rb) \Bigg] \\[2ex]
    &= - \gamma^2 \Bigg[ 2\gamma_{\rm E}^2 \delta(\vec k-\vec k') + 2\gamma_{\rm E} \frac{\delta(k^2-{k'}^2)}{2\pi k} \Pf \lb( \sin^2\tfrac{\theta}{2} \rb)^{-1} \\
    &\qquad\qquad + \frac{\delta(k^2-{k'}^2)}{2\pi k} \Pf \lb( \frac{ \ln\sin^2\tfrac{\theta}{2} }{ \sin^2\tfrac{\theta}{2} } \rb) \Bigg] \\
\label{eq:S2-dist}
\ee
Here, as before, the limit \(\epsilon \to 0^+\) defines the distributions, that is, the limit is taken after integrating with a suitable test functions. We note here that since \(\gamma = \alpha m / k\) each term introduces additional factors of \(1/k\). Thus, these distributions are well-defined on wavefunctions \(\psi(\vec k)\) which vanish in a neighbourhood of \(k=0\). However, such wavefunctions are dense in \(\Hilb\) (see \cite{Dollard:1964cok}) and Herbst \cite{Herbst:1974hi} has shown that the full S-matrix can be extended uniquely to \(\Hilb\) as a unitary operator.

%%----------------------------------------------------------
\section{Computation of the second Born term}
\label{sec:Secondborn}

In this appendix we compute the second Born term required in \cref{sec:second-order}. Let
\be
    I(\epsilon) &= \int d^3\ell~ \frac{1}{ [(\vec p_2-\vec\ell)^2]^{1-\epsilon} [(\vec\ell-\vec p_1)^2]^{1-\epsilon} (\ell^2-p^2-i0^+)} 
\ee
with \(\vec p_1^{\,2}=\vec p_2^{\,2}= p^2\). Writing $\theta$ for the angle between $\vec p_1$ and $\vec p_2$, we define
\be
q^2\equiv(\vec p_2-\vec p_1)^2
=2p^2(1-\cos\theta)
=4p^2\sin^2\frac{\theta}{2}.
\label{eq:qdef-secondborn}
\ee
We first combine the last two denominators using a Feynman parameter \(x\)
\be
\frac{1}{A^{1-\epsilon}B}
=
\frac{\Gamma(2-\epsilon)}{\Gamma(1-\epsilon)}
\int_0^1 dx\,x^{-\epsilon}
\bigl[xA+(1-x)B\bigr]^{-2+\epsilon} \,,
\label{eq:first-feynman}
\ee
with
\be
A=(\vec\ell-\vec p_1)^2,
\qquad
B=\ell^2-p^2-i0^+ \,.
\ee
After shifting $\vec k=\vec\ell-x\vec p_1$, the combined denominator becomes
\be
xA+(1-x)B=k^2-(1-x)^2p^2-i0^+.
\ee
The remaining potential denominator is $(\vec k-\vec P)^2$, where
\be
\vec P\equiv\vec p_2-x\vec p_1,
\qquad
P^2=p^2(1-2x\cos\theta+x^2).
\label{eq:Pdef-secondborn}
\ee
Shifting the integration variable to move the potential denominator to the origin, and using
\be
P^2-(1-x)^2p^2=xq^2,
\ee
we obtain
\be
I(\epsilon)=
\frac{\Gamma(2-\epsilon)}{\Gamma(1-\epsilon)}
\int_0^1 dx\,x^{-\epsilon}
\int d^3k\,
\frac{1}{
(k^2)^{1-\epsilon}
[k^2+2\vec k\cdot\vec P+xq^2-i0^+]^{2-\epsilon}
}.
\label{eq:after-first-feynman}
\ee
Applying Feynman parametrization once more, we find
\be
I(\epsilon)
={}&
\frac{\Gamma(3-2\epsilon)}{\Gamma(1-\epsilon)^2}
\int_0^1 dx\,x^{-\epsilon}
\int_0^1 dy\,y^{-\epsilon}(1-y)^{1-\epsilon} \\
&\times
\int d^3k\,
\lb[
k^2+2(1-y)\vec k\cdot\vec P
+(1-y)xq^2-i0^+
\rb]^{-3+2\epsilon}.
\label{eq:second-feynman}
\ee
Completing the square with $\vec u=\vec k+(1-y)\vec P$ yields
\be
\Delta(x,y)=(1-y)\bigl[xq^2-(1-y)P^2\bigr].
\label{eq:Delta-secondborn}
\ee
We now use the standard momentum integral \cite{Schwartz:2014sze}
\be
\int d^3u\,(u^2+\Delta-i0^+)^{-n}
=
\pi^{3/2}
\frac{\Gamma(n-\frac32)}{\Gamma(n)}
(\Delta-i0^+)^{\frac32-n},
\label{eq:momentum-integral-secondborn}
\ee
initially defined where the integral converges and then analytically continued in the exponent. Taking $n=3-2\epsilon$ in \cref{eq:second-feynman} gives
\be
I(\epsilon)=
\pi^{3/2}
\frac{\Gamma(\frac32-2\epsilon)}{\Gamma(1-\epsilon)^2}
\int_0^1 dx\,x^{-\epsilon}
\int_0^1 dy\,y^{-\epsilon}(1-y)^{1-\epsilon}
(\Delta-i0^+)^{-3/2+2\epsilon}.
\label{eq:afterk}
\ee
To simplify the remaining parameter integral, introduce
\be
z_x\equiv\frac{P^2}{xq^2}
=\frac{1-2x\cos\theta+x^2}{2x(1-\cos\theta)},
\qquad
z_x-1=\frac{(1-x)^2}{2x(1-\cos\theta)}\geq0.
\label{eq:zx-secondborn}
\ee
After the substitution $y\mapsto1-y$, the combination in \cref{eq:Delta-secondborn} takes the form
\be
\Delta-i0^+
=xq^2y\bigl[1-(z_x+i0^+)y\bigr].
\label{eq:Delta-hypergeom}
\ee
The infinitesimal imaginary part therefore fixes the analytic continuation of the hypergeometric function along its branch cut. \Cref{eq:afterk} becomes
\be
I(\epsilon) &= \pi^{\nfrac32} \frac{\Gamma(\frac32-2\epsilon)}{\Gamma(1-\epsilon)^2}
q^{-3+4\epsilon} \int_0^1 dx\,x^{-3/2+\epsilon} \int_0^1dy\,
y^{-1/2+\epsilon}(1-y)^{-\epsilon} [1 - (z_x+i0^+)y]^{-3/2+2\epsilon}. \\
    &= \frac{\pi^{\nfrac32} (2p)^{4\epsilon} }{8p^3} \frac{\Gamma(\frac32-2\epsilon)}{\Gamma(1-\epsilon)^2}
(u)^{-\nfrac32 +2\epsilon} \int_0^1 dx\,x^{-3/2+\epsilon} \int_0^1dy\,
y^{-1/2+\epsilon}(1-y)^{-\epsilon} [1 - (z_x+i0^+)y]^{-3/2+2\epsilon}.
\label{eq:afterk-hypergeom2}
\ee
where
\be
    u = \sin^2\tfrac\theta2 \eqsp z_x = 1 + \frac{(1-x)^2}{4 x u}
\ee
Its contribution to the standard second Born amplitude is
\be
   \braket{ p_2| \op S^{(2)}_{\epsilon, \rm VV}| p_1 } &= +i \frac{4m^2}{(2\pi)^2} \delta(p_2^2 - p_1^2) \frac{1}{2\pi} 4 \alpha^2 \lambda^{4\epsilon} \Gamma^2(1-2\epsilon)\cos^2(\epsilon\pi)~ I(\epsilon) \\
    &= - \gamma^2 \Gamma^2(1-2\epsilon)\cos^2(\epsilon\pi)  \frac{(2p\lambda)^{4\epsilon}}{\epsilon}  \frac{\delta(p_2^2 - p_1^2) }{ 2 \pi p} u^{-1 +2\epsilon} K(\epsilon)
\ee
where
\be
    K(\epsilon) = \epsilon \frac{\Gamma(\frac32-2\epsilon)}{\Gamma(1-\epsilon)^2} \frac{u^{-\half}}{2i\pi^\half} \int_0^1 dx~ x^{-\nfrac32+\epsilon} \int_0^1dy~ y^{-\half+\epsilon}(1-y)^{-\epsilon} [1 - (z_x+i0^+)y]^{-\nfrac32+2\epsilon} 
\ee
We now compute \(K(\epsilon)\) as Laurent series in small \(\epsilon\). It will be convenient to use the Mellin-Barnes representation (see \cite{Smirnov}) for \([1 - (z_x+i0^+)y]^{-\nfrac32+2\epsilon} \) as
\be
    (X+Y)^{-\lambda} = \frac{1}{2\pi i}\int_C dw~ \frac{\Gamma(-w)\Gamma(\lambda+w)}{\Gamma(\lambda)} X^{-\lambda-w} Y^{w} \,,
\ee
with $\lambda=\tfrac32-2\epsilon$, $X=1-y$ and $Y=y(1-z_x-i0^+)$. The contour \(C\) is chosen with \(\Re w = \text{constant}\), i.e., an infinite vertical line in the complex \(w\)-plane, so that the poles of \(\Gamma(\ldots + w)\) are to the left and those of \(\Gamma(\ldots - w)\) are to the right. Substituting into the expression for \(K(\epsilon)\), the integrals over $x$ and $y$ can be done in terms of Gamma functions to get
\be
    K(\epsilon) &= \frac{\epsilon u^{-\half}}{2i\pi^\half \Gamma(1-\epsilon)^2\Gamma(2\epsilon)} \\
    &\qquad \times \int_C \frac{dw}{2\pi i}~ \Gamma(-w) \Gamma\lb(\nfrac32 - 2\epsilon + w\rb) \Gamma\lb(-\half + \epsilon -w \rb)^2 \Gamma(1 + 2w) e^{-i\pi w}(4u)^{-w}.
\ee
The contour \(C\) lies in $-\tfrac12<\mathrm{Re}\,w<-\tfrac12+\epsilon$. To evaluate this integral, we note that, $u^{-w}$ decays as $\Re w \to -\infty$ (since $0<u<1$). So we close the contour to the left, counterclockwise and use the residue theorem. The (simple) poles enclosed are $w_n=-\tfrac{n+1}{2}$ from $\Gamma(1+2w)$ and $w'_n=-\nfrac32+2\epsilon-n$ from $\Gamma\lb(\nfrac32-2\epsilon+w\rb)$, with $n \geq 0$, and the relevant residues are
\be
    \mathop{\rm Res}\limits_{w = w_n} \Gamma(1+2w) = \frac{(-1)^n}{2 n! } \eqsp \mathop{\rm Res}\limits_{w = w_n'} \Gamma\lb(\nfrac32-2\epsilon+w\rb) = \frac{(-1)^n}{n!} \,.
\ee
This gives
\be
    K(\epsilon) &= \frac{\epsilon}{ \Gamma(1-\epsilon)^2\Gamma(2\epsilon)} \sum_{n=0}^\infty \lb[ A_n + B_n \rb] = \frac{ 2 \epsilon^2}{ \Gamma(1-\epsilon)^2\Gamma(1+2\epsilon)} \sum_{n=0}^\infty \lb[ A_n + B_n \rb]
\ee
where
\be
    A_n &= \frac{1}{2\pi^\half} \frac{(-2i)^n}{n!}\Gamma\lb(\tfrac{n+1}{2}\rb) \Gamma\lb(1-2\epsilon-\tfrac n2\rb) \Gamma\lb(\epsilon+\tfrac n2\rb)^2~ u^{n/2} \\[2ex]
    B_n &= - e^{-2\pi i\epsilon} \frac{ 2^{2-4\epsilon}}{\pi^\half} \frac{4^n}{n!} \Gamma\lb(\tfrac32-2\epsilon+n\rb) \Gamma(1-\epsilon+n)^2 \Gamma(-2+4\epsilon-2n) u^{n + 1-2\epsilon} \,.
\ee
We now extract the relevant Laurent series for \(K(\epsilon)\). The \(A_0\) is the only term which has a double pole \(\sim 1/\epsilon^2\) and is constant in \(u\). This term gives the \(O(1)\) term in \(K\) given by
\be
    K_0 &\sim \frac{ 2 \epsilon^2}{ \Gamma(1-\epsilon)^2\Gamma(1+2\epsilon)} A_0 = \frac{ 2 \epsilon^2}{ \Gamma(1-\epsilon)^2\Gamma(1+2\epsilon)} \frac{1}{2\pi^\half} \Gamma\lb(\tfrac{1}{2}\rb) \Gamma\lb(1-2\epsilon \rb) \Gamma\lb(\epsilon\rb)^2 \\
    &= 1 + O(\epsilon^3)
\label{eq:K0}
\ee
The terms \(A_n\) with even \(n = 2k\) and all the \(B_n\) terms with \(n = k-1\) contain a simple pole \(\sim 1/\epsilon\), with the series expansion
\be
    A_{2k} &= \frac{1}{2\pi^\half} \frac{(-2i)^{2k}}{(2k)!}\Gamma\lb(k + \half \rb) \Gamma\lb(1-2\epsilon-k \rb) \Gamma\lb(\epsilon + k \rb)^2~ u^{k} \\
    &= \frac{ \pi}{ 2 k! \sin(2\pi\epsilon)} \frac{ \Gamma\lb(\epsilon + k \rb)^2}{ \Gamma(k+2\epsilon) }~ u^{k} = \frac{u^k}{4 k \epsilon} + O(\epsilon) \\[2ex]
    B_{k-1} &= - e^{-2\pi i\epsilon} \frac{ 2^{2-4\epsilon}}{\pi^\half} \frac{4^{k-1}}{(k-1)!} \Gamma\lb(\half - 2\epsilon + k \rb) \Gamma(-\epsilon + k)^2 \Gamma(4\epsilon - 2k) u^{k - 2\epsilon} \\
    &= - \frac{u^k}{4k} \lb[ \frac{1}{\epsilon} + \frac{2}{k} - 2\pi i - 2 \ln u \rb] + O(\epsilon)
\ee
Thus, the pole cancels when these are summed and we get
\be
    \sum_{k = 1}^\infty \lb[ A_{2k} +  B_{k-1} \rb] &= - \sum_{k = 1}^\infty \frac{u^k}{4k} \lb[  \frac{2}{k} - 2\pi i - 2 \ln u \rb] + O(\epsilon) \\
    &= - \tfrac12 \lb[ \text{Li}_2(u) + (i \pi + \ln u) \ln (1-u) \rb] + O(\epsilon)
\label{eq:A-even-B}
\ee
Finally, we have the \(A_n\) for odd \(n = 2k+1\), where each term is finite at \(\epsilon = 0\) and we get
\be
    \sum_{k = 0}^\infty A_{2k+1} &= \sum_{k = 0}^\infty \frac{1}{2\pi^\half} \frac{(-2i)^{(2k+1)}}{(2k+1)!} \Gamma\lb(k+1\rb) \Gamma\lb(\half - k \rb) \Gamma\lb(\half + k \rb)^2~ u^{k + \half} + O(\epsilon) \\
    &= - \sum_{k = 0}^\infty \frac{i \pi }{2k+1} u^{k + \half} + O(\epsilon) = - \frac{i \pi}{2} \ln \frac{1 + u^\half}{1 - u^\half} + O(\epsilon)
\label{eq:A-odd}
\ee
Using $\ln(1-u)=\ln(1- u^\half)+\ln(1+ u^\half)$, the sum of \cref{eq:A-even-B,eq:A-odd} gives the \(O(\epsilon^2)\) term of \(K\)
\be
    K_2 \sim - \epsilon^2 \bigg[ \text{Li}_2(u) + 2 \pi i \ln (1+u^\half) + \ln u \ln (1-u) \bigg] + O(\epsilon^3)
\label{eq:K1}
\ee
Thus, from \cref{eq:K0,eq:K1} we get
\be
    K &= 1 - \epsilon^2 R(u) + O(\epsilon^3) \\
    \text{where } R(u) &= \text{Li}_2(u) + 2 \pi i \ln (1+u^\half) + \ln u \ln (1-u) \,,
\label{eq:K-fin}
\ee
and we note that \(R(u=0) = 0\). Thus,
\be
 \braket{ p_2| \op S^{(2)}_{\epsilon, \rm VV}| p_1 }
    &= - \gamma^2 \Gamma^2(1-2\epsilon)\cos^2(\epsilon\pi)  \frac{(2p\lambda)^{4\epsilon}}{\epsilon}  \frac{\delta(p_2^2 - p_1^2) }{ 2 \pi p} u^{-1 +2\epsilon} \lb[ 1 - \epsilon^2 R(u) + O(\epsilon^3)\rb]
\ee

%%THE END

\bibliographystyle{JHEP}
\bibliography{sample}      

@article{Dalitz:1951ah,
    author = "Dalitz, R. H.",
    title = "{On higher Born approximations in potential scattering}",
    doi = "10.1098/rspa.1951.0085",
    journal = "Proc. Roy. Soc. Lond. A",
    volume = "206",
    pages = "509--520",
    year = "1951"
}

@article{Dollard:1964cok,
    author = "Dollard, John D.",
    title = "{Asymptotic Convergence and the Coulomb Interaction}",
    doi = "10.1063/1.1704171",
    journal = "J. Math. Phys.",
    volume = "5",
    number = "6",
    pages = "729",
    year = "1964"
}

@article{Herbst:1974hi,
    author = "Herbst, I. W.",
    title = "{On the connectedness structure of the Coulomb S-matrix}",
    doi = "10.1007/BF01646192",
    journal = "Commun. Math. Phys.",
    volume = "35",
    pages = "181--191",
    year = "1974"
}

@article{Guth-Mullin,
    author = "Guth, E. and Mullin, C.J.",
    title = "{Momentum representation of the Coulomb scattering wave functions}",
    doi = "10.1103/PhysRev.83.667",
    journal = "Phys. Rev.",
    volume = "83",
    pages = "667",
    year = "1951"
}

@book{Reed-Simon3,
author = "Reed, M. and Simon, B.",
title = "{Scattering Theory}",
series = "Methods of Modern Mathematical Physics",
volume = "III",
year = "{1979}",
publisher = "Academic Press",
address ="San Diego, London"
}

@article{Lippstreu:2023vvg,
    author = "Lippstreu, Luke",
    title = "{A perturbation theory for the Coulomb-phase infrared divergence}",
    eprint = "2312.08455",
    archivePrefix = "arXiv",
    primaryClass = "hep-th",
    month = "12",
    year = "2023"
}

@article{Lippstreu:2025jit,
    author = "Lippstreu, Luke",
    title = "{Analytic Properties of Infrared-Finite Amplitudes in Theories with Long-Range Forces}",
    eprint = "2505.04702",
    archivePrefix = "arXiv",
    primaryClass = "hep-th",
    month = "5",
    year = "2025"
}

@article{Kulish:1970ut,
    author = "Kulish, P. P. and Faddeev, L. D.",
    title = "{Asymptotic conditions and infrared divergences in quantum electrodynamics}",
    reportNumber = "D70-07927",
    doi = "10.1007/BF01066485",
    journal = "Theor. Math. Phys.",
    volume = "4",
    pages = "745",
    year = "1970"
}

@article{Schwinger:1964zzb,
    author = "Schwinger, Julian",
    title = "{Coulomb Green's Function}",
    doi = "10.1063/1.1931195",
    journal = "J. Math. Phys.",
    volume = "5",
    pages = "1606--1608",
    year = "1964"
}

@book{Landau:Lifschitz,
author = "L. D. Landau and E. M. Lifshitz", 
title = "Quantum Mechanics: Non-Relativistic Theory, vol. v.3 of
Course of Theoretical Physics",
publisher = "Butterworth-Heinemann",
address = "Oxford",
year = "1991"
}

@article{Morchio:2014zga,
    author = "Morchio, G. and Strocchi, F.",
    title = "{Dynamics of Dollard asymptotic variables. Asymptotic fields in Coulomb scattering}",
    eprint = "1410.5612",
    archivePrefix = "arXiv",
    primaryClass = "math-ph",
    doi = "10.1142/S0129055X1650001X",
    journal = "Rev. Math. Phys.",
    volume = "28",
    number = "01",
    pages = "1650001",
    year = "2016"
}

@article{Taylor:1973da,
    author = "Taylor, John R.",
    title = "{A new rigorous approach to Coulomb scattering}",
    reportNumber = "ICTP-73-7",
    doi = "10.1007/BF02723639",
    journal = "Nuovo Cim. B",
    volume = "23",
    pages = "313--334",
    year = "1974"
}

@article{Prabhu:2022zcr,
    author = "Prabhu, Kartik and Satishchandran, Gautam and Wald, Robert M.",
    title = "{Infrared finite scattering theory in quantum field theory and quantum gravity}",
    eprint = "2203.14334",
    archivePrefix = "arXiv",
    primaryClass = "hep-th",
    doi = "10.1103/PhysRevD.106.066005",
    journal = "Phys. Rev. D",
    volume = "106",
    number = "6",
    pages = "066005",
    year = "2022"
}

@article{Prabhu:2024zwl,
    author = "Prabhu, Kartik and Satishchandran, Gautam",
    title = "{Infrared finite scattering theory: scattering states and representations of the BMS group}",
    eprint = "2402.00102",
    archivePrefix = "arXiv",
    primaryClass = "hep-th",
    doi = "10.1007/JHEP08(2024)055",
    journal = "JHEP",
    volume = "08",
    pages = "055",
    year = "2024"
}

@article{Prabhu:2024lmg,
    author = "Prabhu, Kartik and Satishchandran, Gautam",
    title = "{Infrared finite scattering theory: Amplitudes and soft theorems}",
    eprint = "2402.18637",
    archivePrefix = "arXiv",
    primaryClass = "hep-th",
    doi = "10.1103/PhysRevD.110.085022",
    journal = "Phys. Rev. D",
    volume = "110",
    number = "8",
    pages = "085022",
    year = "2024"
}

@article{Weinberg:1965nx,
    author = "Weinberg, Steven",
    title = "{Infrared photons and gravitons}",
    doi = "10.1103/PhysRev.140.B516",
    journal = "Phys. Rev.",
    volume = "140",
    pages = "B516--B524",
    year = "1965"
}

@book{Ashtekar:1987tt,
      author         = "Ashtekar, A.",
      title          = "{Asymptotic Quantization: Based On 1984 Naples Lectures}",
    publisher = "{Bibliopolis}",
    address = "Naples, Italy",
    year = 1987,
      series        = "Monographs and Textbooks in Physical Science",
      SLACcitation   = "%%CITATION = INSPIRE-256903;%%"
}

@article{Kapec:2017tkm,
    author = "Kapec, Daniel and Perry, Malcolm and Raclariu, Ana-Maria and Strominger, Andrew",
    title = "{Infrared Divergences in QED, Revisited}",
    eprint = "1705.04311",
    archivePrefix = "arXiv",
    primaryClass = "hep-th",
    doi = "10.1103/PhysRevD.96.085002",
    journal = "Phys. Rev. D",
    volume = "96",
    number = "8",
    pages = "085002",
    year = "2017"
}

@article{Choi:2017ylo,
    author = "Choi, Sangmin and Akhoury, Ratindranath",
    title = "{BMS Supertranslation Symmetry Implies Faddeev-Kulish Amplitudes}",
    eprint = "1712.04551",
    archivePrefix = "arXiv",
    primaryClass = "hep-th",
    doi = "10.1007/JHEP02(2018)171",
    journal = "JHEP",
    volume = "02",
    pages = "171",
    year = "2018"
}

@article{Gabai:2016kuf,
    author = "Gabai, Barak and Sever, Amit",
    title = "{Large gauge symmetries and asymptotic states in QED}",
    eprint = "1607.08599",
    archivePrefix = "arXiv",
    primaryClass = "hep-th",
    doi = "10.1007/JHEP12(2016)095",
    journal = "JHEP",
    volume = "12",
    pages = "095",
    year = "2016"
}

@article{Lehmann:1954rq,
    author = "Lehmann, H. and Symanzik, K. and Zimmermann, W.",
    title = "{Zur Formulierung quantisierter Feldtheorien}",
    doi = "10.1007/BF02731765",
    journal = "Nuovo Cim.",
    volume = "1",
    pages = "205--225",
    year = "1955",
    note    = {English translation: {\it On the formulation of quantized field theories} in {\it Dispersion Relations and the Abstract Approach to Field Theory}, ed. L. Klein, Gordon and Breach, New York, 1961}
}

@article{Kibble:1968sfb,
    author = "Kibble, T. W. B.",
    title = "{Coherent Soft-Photon States and Infrared Divergences. I. Classical Currents}",
    doi = "10.1063/1.1664582",
    journal = "J. Math. Phys.",
    volume = "9",
    number = "2",
    pages = "315--324",
    year = "1968"
}

@article{Kibble:1968lka,
    author = "Kibble, T. W. B.",
    title = "{Coherent soft-photon states and infrared divergences. IV. The scattering operator}",
    doi = "10.1103/PhysRev.175.1624",
    journal = "Phys. Rev.",
    volume = "175",
    pages = "1624--1640",
    year = "1968"
}

@article{Kibble:1968npb,
    author = "Kibble, T. W. B.",
    title = "{Coherent soft-photon states and infrared divergences. III. Asymptotic states and reduction formulas}",
    doi = "10.1103/PhysRev.174.1882",
    journal = "Phys. Rev.",
    volume = "174",
    pages = "1882--1901",
    year = "1968"
}

@article{Kibble:1968oug,
    author = "Kibble, T. W. B.",
    title = "{Coherent soft-photon states and infrared divergences. II. Mass-shell singularities of Green's functions}",
    doi = "10.1103/PhysRev.173.1527",
    journal = "Phys. Rev.",
    volume = "173",
    pages = "1527--1535",
    year = "1968"
}

@article{Chung:1965zza,
    author = "Chung, Victor",
    title = "{Infrared Divergence in Quantum Electrodynamics}",
    doi = "10.1103/PhysRev.140.B1110",
    journal = "Phys. Rev.",
    volume = "140",
    pages = "B1110--B1122",
    year = "1965"
}

@article{Dybalski:2017mip,
    author = "Dybalski, Wojciech",
    title = "{From Faddeev{\textendash}Kulish to LSZ. Towards a non-perturbative description of colliding electrons}",
    eprint = "1706.09057",
    archivePrefix = "arXiv",
    primaryClass = "hep-th",
    doi = "10.1016/j.nuclphysb.2017.10.018",
    journal = "Nucl. Phys. B",
    volume = "925",
    pages = "455--469",
    year = "2017"
}

@article{Duch:2019wpf,
    author = "Duch, Pawe{\l}",
    title = "{Infrared problem in perturbative quantum field theory}",
    eprint = "1906.00940",
    archivePrefix = "arXiv",
    primaryClass = "math-ph",
    doi = "10.1142/S0129055X2150032X",
    journal = "Rev. Math. Phys.",
    volume = "33",
    number = "10",
    pages = "2150032",
    year = "2021"
}

@article{Papanicolaou:1976sv,
    author = "Papanicolaou, N.",
    title = "{Infrared Problems in Quantum Electrodynamics}",
    doi = "10.1016/0370-1573(76)90003-X",
    journal = "Phys. Rept.",
    volume = "24",
    pages = "229--313",
    year = "1976"
}

@article{Zwanziger:1973if,
    author = "Zwanziger, D.",
    title = "{Reduction formulas for charged particles and coherent states in quantum electrodynamics}",
    doi = "10.1103/PhysRevD.7.1082",
    journal = "Phys. Rev. D",
    volume = "7",
    pages = "1082--1099",
    year = "1973"
}

@article{Forde:2003jt,
    author = "Forde, Darren A. and Signer, Adrian",
    title = "{Infrared finite amplitudes for massless gauge theories}",
    eprint = "hep-ph/0311059",
    archivePrefix = "arXiv",
    reportNumber = "IPPP-03-67, DCPT-03-134",
    doi = "10.1016/j.nuclphysb.2004.02.024",
    journal = "Nucl. Phys. B",
    volume = "684",
    pages = "125--161",
    year = "2004"
}

@article{Hannesdottir:2019opa,
    author = "Hannesdottir, Holmfridur and Schwartz, Matthew D.",
    title = "{$S$ -Matrix for massless particles}",
    eprint = "1911.06821",
    archivePrefix = "arXiv",
    primaryClass = "hep-th",
    doi = "10.1103/PhysRevD.101.105001",
    journal = "Phys. Rev. D",
    volume = "101",
    number = "10",
    pages = "105001",
    year = "2020"
}

@book{Landau:1975pou,
    author = "Landau, L. D. and Lifschitz, E. M.",
    title = "{Mechanics}",
    publisher = "Pergamon Press",
    address = "Oxford",
    series = "Course of Theoretical Physics",
    volume = "1",
    year = "1976"
}

@book{Schwartz:2014sze,
    author = "Schwartz, Matthew D.",
    title = "{Quantum Field Theory and the Standard Model}",
    isbn = "978-1-107-03473-0, 978-1-107-03473-0",
    publisher = "Cambridge University Press",
    month = "3",
    year = "2014"
}

@book{Arnold:1989who,
    author = "Arnold, V. I.",
    title = "{Mathematical Methods of Classical Mechanics}",
    publisher = "Springer",
    series = "Graduate Texts in Mathematics",
    volume = "60",
    year = "1989"
}

@book{Goldstein:1950,
    author = "Goldstein, H. and Poole, C. and Safko, J.",
    title = "Classical Mechanics",
    publisher = "Addison-Wesley",
    year = "1950"
}

@book{Gelfand1,
      author         = "Gel'fand, I. M. and Shilov, G. E.",
      title          = "{Properties and Operations}",
      series = "Generalized Functions",
    volume = "1",
      year           = "1964",
      publisher      = "Academic Press",
      address = "New York and London",
      note = "Translated by: E. Saletan"
}

@book{Smirnov,
    author = "Smirnov, Vladimir A.",
    title = "{Analytic tools for Feynman integrals}",
    series = "Springer Tracts in Modern Physics",
    volume = "250",
    publisher = "Springer Berlin",
    year = "2012"
}

@article{Lippstreu:2026xfw,
    author = "Lippstreu, Luke",
    title = "{Unitarity and the Forward Direction in Theories with Long-Range Forces}",
    eprint = "2609.16896",
    archivePrefix = "arXiv",
    primaryClass = "hep-th",
    month = "9",
    year = "2026"
}

@article{Ashtekar:2018lor,
    author = "Ashtekar, Abhay and Campiglia, Miguel and Laddha, Alok",
    title = "{Null infinity, the BMS group and infrared issues}",
    eprint = "1808.07093",
    archivePrefix = "arXiv",
    primaryClass = "gr-qc",
    doi = "10.1007/s10714-018-2464-3",
    journal = "Gen. Rel. Grav.",
    volume = "50",
    number = "11",
    pages = "140--163",
    year = "2018"
}

@article{Choi:2024mac,
    author = "Choi, Sangmin and Laddha, Alok and Puhm, Andrea",
    title = "{The classical super-phaserotation infrared triangle. Classical logarithmic soft theorem as conservation law in (scalar) QED}",
    eprint = "2412.16149",
    archivePrefix = "arXiv",
    primaryClass = "hep-th",
    doi = "10.1007/JHEP05(2025)155",
    journal = "JHEP",
    volume = "05",
    pages = "155",
    year = "2025"
}

@article{Dollard-review,
    author = "Dollard, J.D.",
    title ="{Quantum-mechanical scattering theory for
short-range and Coulomb interactions}" ,
    journal = "Rocky Mountain J. Math.",
    volume = 1,
    number = 1,
    year = "1971"
}

@article{Campiglia:2015qka,
    author = "Campiglia, Miguel and Laddha, Alok",
    title = "{Asymptotic symmetries of QED and Weinberg{\textquoteright}s soft photon theorem}",
    eprint = "1505.05346",
    archivePrefix = "arXiv",
    primaryClass = "hep-th",
    doi = "10.1007/JHEP07(2015)115",
    journal = "JHEP",
    volume = "07",
    pages = "115",
    year = "2015"
}

@article{Campiglia:2019wxe,
    author = "Campiglia, Miguel and Laddha, Alok",
    title = "{Loop Corrected Soft Photon Theorem as a Ward Identity}",
    eprint = "1903.09133",
    archivePrefix = "arXiv",
    primaryClass = "hep-th",
    doi = "10.1007/JHEP10(2019)287",
    journal = "JHEP",
    volume = "10",
    pages = "287",
    year = "2019"
}

@article{Choi:2024ajz,
    author = "Choi, Sangmin and Laddha, Alok and Puhm, Andrea",
    title = "{The classical super-rotation infrared triangle. Classical logarithmic soft theorem as conservation law in gravity}",
    eprint = "2412.16142",
    archivePrefix = "arXiv",
    primaryClass = "hep-th",
    doi = "10.1007/JHEP04(2025)138",
    journal = "JHEP",
    volume = "04",
    pages = "138",
    year = "2025"
}

@article{Kim:2023qbl,
    author = "Kim, Seolhwa and Kraus, Per and Monten, Ruben and Myers, Richard M.",
    title = "{S-matrix path integral approach to symmetries and soft theorems}",
    eprint = "2307.12368",
    archivePrefix = "arXiv",
    primaryClass = "hep-th",
    doi = "10.1007/JHEP10(2023)036",
    journal = "JHEP",
    volume = "10",
    pages = "036",
    year = "2023"
}

@article{Isen:2026xoc,
    author = "Isen, Jack and Kraus, Per and Monten, Ruben and Myers, Richard M.",
    title = "{The gravitational S-matrix from the path integral: asymptotic symmetries and soft theorems}",
    eprint = "2603.17045",
    archivePrefix = "arXiv",
    primaryClass = "hep-th",
    reportNumber = "CERN-TH-2026-044",
    month = "3",
    year = "2026"
}

@article{Arefeva:1974jv,
    author = "Arefeva, I. Ya. and Faddeev, L. D. and Slavnov, A. A.",
    title = "{Generating Functional for the s Matrix in Gauge Theories}",
    reportNumber = "SACLAY-DPH-T-74-44",
    doi = "10.1007/BF01038094",
    journal = "Teor. Mat. Fiz.",
    volume = "21",
    pages = "311--321",
    year = "1974"
}

@article{Ammon:2026svb,
    author = "Ammon, Martin and Brandts, Konrad and Capone, Federico and Hollweck, Jakob",
    title = "{A generating functional for infrared-safe QED amplitudes}",
    eprint = "2608.18205",
    archivePrefix = "arXiv",
    primaryClass = "hep-th",
    month = "8",
    year = "2026"
}

@article{Choi:2019fuq,
    author = "Choi, Sangmin and Sandeep Pradhan, S. and Akhoury, Ratindranath",
    title = "{Supertranslation Hair of Schwarzschild Black Hole: A Wilson Line Perspective}",
    eprint = "1910.05882",
    archivePrefix = "arXiv",
    primaryClass = "hep-th",
    doi = "10.1007/JHEP01(2020)013",
    journal = "JHEP",
    volume = "01",
    pages = "013",
    year = "2020"
}

@article{Jain:2023fxc,
    author = "Jain, Diksha and Kundu, Suman and Minwalla, Shiraz and Parrikar, Onkar and Prabhu, Siddharth G. and Shrivastava, Pushkal",
    title = "{The S-matrix and boundary correlators in flat space}",
    eprint = "2311.03443",
    archivePrefix = "arXiv",
    primaryClass = "hep-th",
    doi = "10.1007/JHEP02(2026)151",
    journal = "JHEP",
    volume = "02",
    pages = "151",
    year = "2026"
}

@article{Fabbrichesi:1993kz,
    author = "Fabbrichesi, M. and Pettorino, Roberto and Veneziano, G. and Vilkovisky, G. A.",
    title = "{Planckian energy scattering and surface terms in the gravitational action}",
    eprint = "hep-th/9309037",
    archivePrefix = "arXiv",
    reportNumber = "CERN-TH-6904-93-REV, CERN-TH-6904-93",
    doi = "10.1016/0550-3213(94)90361-1",
    journal = "Nucl. Phys. B",
    volume = "419",
    pages = "147--188",
    year = "1994"
}

@article{Upadhyay:2025ged,
    author = "Upadhyay, Shivam",
    title = "{On symmetries of gravitational on-shell boundary action at null infinity}",
    eprint = "2501.07136",
    archivePrefix = "arXiv",
    primaryClass = "hep-th",
    month = "1",
    year = "2025"
}

@article{Kacser-C,
    author = "Kacser, C.",
    title = "{Higher born approximations in non-relativistic Coulomb scattering}",
    journal = "Nuovo Cim." ,
    doi="10.1007/BF02732939",
    year = "1959"
}

@article{Bander:1965im,
    author = "Bander, Myron and Itzykson, Claude",
    title = "{Group Theory and the Hydrogen Atom. II.}",
    reportNumber = "SLAC-PUB-0141",
    doi = "10.1103/RevModPhys.38.346",
    journal = "Rev. Mod. Phys.",
    volume = "38",
    pages = "346--358",
    year = "1966"
}

@article{Zwanziger1967,
    author = {Zwanziger, Daniel},
    title = "{Algebraic Calculation of Nonrelativistic Coulomb Phase Shifts}",
    journal = {J. Math. Phys.},
    volume = {8},
    number = {9},
    pages = {1858-1860},
    year = {1967},
    month = {09},
    issn = {0022-2488},
    doi = {10.1063/1.1705429},
    url = {https://doi.org/10.1063/1.1705429}
}

@article{Nekrasov:2023kxu,
    author = "Nekrasov, M. L.",
    title = "{Coulomb-nuclear interference in elastic proton scattering in the eikonal approach}",
    eprint = "2311.09330",
    archivePrefix = "arXiv",
    primaryClass = "hep-ph",
    doi = "10.1016/j.physletb.2024.138485",
    journal = "Phys. Lett. B",
    volume = "849",
    pages = "138485",
    year = "2024"
}

@article{Kabat:1992tb,
    author = "Kabat, Daniel N. and Ortiz, Miguel",
    title = "{Eikonal quantum gravity and Planckian scattering}",
    eprint = "hep-th/9203082",
    archivePrefix = "arXiv",
    reportNumber = "MIT-CTP-2069",
    doi = "10.1016/0550-3213(92)90627-N",
    journal = "Nucl. Phys. B",
    volume = "388",
    pages = "570--592",
    year = "1992"
}
\end{document}